\documentclass[smallextended,envcountsect]{svjour3}
\pdfoutput=1
\usepackage{xspace}
\usepackage{xcolor}

\newcommand{\subject}{\emph{subject}\xspace}
\newcommand{\corpus}{\emph{corpus}\xspace}
\newcommand{\candidate}{\emph{candidate}\xspace}
\newcommand{\candidates}{\emph{candidates}\xspace}
\newcommand{\pypi}{PyPI\xspace}
\newcommand{\package}[1]{\emph{#1}\xspace}
\newcommand{\code}[1]{\texttt{#1}\xspace}

\newcommand{\rqonetext}{How frequent are 
  global identifiers among open source Python packages?}
\newcommand{\rqtwotext}{How many global identifiers are needed to \emph{narrow down}
  the origin of a given source code artifact to a small set of candidates,
  within the open source Python ecosystem?}

\newcommand{\fingerp}{\emph{fingerprint}\xspace}
\newcommand{\bl}{\emph{blocklist}\xspace}

\newcommand{\validationDescription}{We validate the proposed approach by mapping Debian source packages
  implemented in Python to the corresponding \pypi packages; this approach uses at most five trials, where each trial 
  uses three randomly chosen global identifiers
  from a randomly chosen python file of the subject software package, then ranks results using a popularity index and requires to inspect only the top result.
  In our experiments, this method is effective at finding the  true origin of a project
  with a recall of 0.9 and precision of 0.77.}

\usepackage{framed}
\usepackage{xparse}
\newsavebox{\fminipagebox}
\NewDocumentEnvironment{hassanbox}{m O{\fboxsep}}
 {\par\kern#2\noindent\begin{lrbox}{\fminipagebox}
  \begin{minipage}{#1}\ignorespaces}
 {\end{minipage}\end{lrbox}\makebox[#1]{\kern\dimexpr-\fboxsep-\fboxrule\relax
    \fbox{\usebox{\fminipagebox}}\kern\dimexpr-\fboxsep-\fboxrule\relax
  }\par\kern#2
 }

 \usepackage{amsmath,amssymb,amsfonts}
\usepackage{enumerate}
\usepackage{lipsum}
\usepackage{graphicx}
\usepackage[colorlinks,urlcolor=black,linkcolor=black,citecolor=black,linktocpage=true]{hyperref}
\usepackage{cleveref}
\usepackage{siunitx}
\usepackage{csvsimple,booktabs,supertabular,longtable}

\begin{document}
\title{Using the Uniqueness of Global Identifiers to Determine the Provenance of Python Software Source Code}

\author{Yiming Sun\and
  Daniel German\and
  Stefano Zacchiroli}

\institute{
  Y.~Sun
  \at University of Victoria, Victoria, Canada
  \email{yimings@uvic.ca}
  \and
  D.~German
  \at University of Victoria, Victoria, Canada
  \email{dmg@uvic.ca}
  \and
  S.~Zacchiroli
  \at LTCI, Télécom Paris, Institut Polytechnique de Paris, Paris, France,\newline
  \email{stefano.zacchiroli@telecom-paris.fr}
  \and
}

 \maketitle
\begin{abstract}
  We consider the problem of identifying the provenance of free/open source
  software (FOSS) and specifically the need of identifying where reused source
  code has been copied from. We propose a
  lightweight approach to solve the problem based on software identifiers---such as the names of
  variables, classes, and functions chosen by programmers. The proposed
  approach is able to efficiently narrow down to a small set of candidate origin
  products, to be further analyzed with more expensive techniques to make a
  final provenance determination.

  By analyzing the \pypi (Python Packaging Index) open source ecosystem we find that globally defined identifiers are very distinct. Across
  \pypi's 244\,K packages we found 11.2\,M different global identifiers (classes and
  method/function names---with only 0.6\% of identifiers shared among the two
  types of entities); 76\% of identifiers were  used only in one package, and 93\% in at most
  3.  Randomly selecting 3 non-frequent global identifiers from an input product is enough to narrow down its origins to a
  maximum of 3 products within 89\% of the cases.

  \validationDescription

\end{abstract}

\section{Introduction}
\label{sec:intro}

In modern software development, applications
are rarely built from scratch. Rather, software is for the most
part~\cite{securosis,synopsys2020ossra} built reusing existing free/open source
software (FOSS) components, mixing in varying amounts of custom in-house code.
A 2014 survey claims that at least 75\% of organizations rely on open source as
the foundation of their applications \cite{securosis}; a 2020
analysis~\cite{synopsys2020ossra} by an industry player in the field of mergers
and acquisitions reports that 99\% of audited code bases contain FOSS
components, with 70\% of all audited code being itself open source software.

In terms of development practices, reuse of open source code happens in
different forms: from retrieving and integrating entire FOSS components (also
known as ``software vendoring''~\cite{zimmermann2020vendoring}), to simply copying chunks of publicly
accessible code and pasting them into the source code files of the software
under development.
While useful for speeding up development and believed to have benefits on code
quality, coding efficiency, and maintenance~\cite{gharehyazie2017some}, FOSS
code reuse requires proper management of the software supply
chain to avoid nefarious side-effects~\cite{harutyunyan2020supplychain}.
From a security perspective, for example, operation engineers will need to
monitor the status of all software components deployed in production and keep
them updated when newer versions that fix security flaws are released.  As a
matter of concern, 88\% of applications audited in the previously mentioned
study contained open source dependencies that underwent no development activity over
the two previous years~\cite{synopsys2020ossra}.

From a legal point of view, when the source code of an application made
available to end-users contains parts copied from other FOSS components, the
software distributor is responsible to ensure that all involved licenses (open
source or otherwise) are mutually compatible and consistent with the
applicable end-user software license~\cite{rosen2005fosslicensing}. Short of
that, the distributor might incur significant legal and financial risks due to
potential copyright violations.

The state-of-the-art approach to minimize both security and legal risks related
to the reuse of FOSS components is based on a set of practices and tools that
are plugged into the software build process~\cite{phipps2020compliance}, e.g.,
as part of continuous integration (CI) pipelines. The software being built is
automatically analyzed to determine what are the main software components it
contains---a process known as Software Composition Analysis, or
SCA~\cite{ombredanne2020sca}---which results in the production of a
Software Bill of Material, or SBOM~\cite{spdx}. Then, licensing and
security information about each identified component are retrieved and verified
for adherence to custom in-house policies, failing the build and triggering
further audits or decisions when necessary.

A key ingredient of these pipelines is the ability to \textbf{identify where
  the source code being built comes from}, that is, \textbf{determine the
  \emph{provenance} of software source code artifacts} at various
granularities: entire source code trees, individual source code files, brief code
snippets. A failure in identifying the provenance of source code can
result in overlooking relevant information about security or licensing issues,
with potentially severe consequences. In addition to this \emph{correctness}
requirement, and due to the need of deploying provenance tracking solutions as
part of automated workflows, lookup \emph{efficiency} is often a key
factor in deciding whether an approach is practically usable or not.

Several techniques are available today to determine the provenance of source
code artifacts. However, searching for the occurrence of an entity in vast
bodies of open source code remains challenging. For instance, applying
conventional clone detection methods such as text-based comparison and AST
matching is computationally expensive and quickly becomes impractical.
Conversely, methods based on exact file matching, e.g., based on cryptographic
hashes, will fail to identify relevant software origins when even very minor
file changes are applied. A satisfactory software provenance identification
technique should be fast, scalable, and capable of finding
matches even in the presence of some code changes, in order to be
practically useful.

\paragraph{Contributions.}

In this paper \textbf{we propose a lightweight approach to determine the
  provenance of source code artifacts based on the uniqueness of
  identifiers}---i.e., the names chosen by developers to reference common
programming abstractions like variables, data types, classes, methods, etc. We
aim to show that, by first \emph{indexing} the set of identifiers found in a
large corpus of open source software components, and then \emph{querying} the index
using as input the identifiers found in a given source component of unknown origin,
it is both practically possible and efficient to determine its
provenance (within the corpus).
Specifically, we consider the software ecosystem consisting of open source
Python packages available from the Python Package Index
(\pypi)\footnote{\url{https://pypi.org/}, accessed 2021-11-15}, for a total of
244 thousand packages as of March 2021.

As foundational empirical evidence to validate the proposed approach we will
first answer the research question:
\begin{enumerate}[\bfseries RQ1]
\item \label{rq:i} \rqonetext
\end{enumerate}

We will answer this question at package level. We define the \emph{frequency} of an identifier as the number of software
products (or ``packages'', according to \pypi terminology) that define such
identifier within a given corpus. More generally we will
characterize the distribution of global identifier popularity.  We find that global identifiers declared by
programmers tend to be unique.
About 76\% of class and function names uniquely identify a software product
among the 244\,K packages in our corpus (have frequency equal to 1); and that up to 93\% of them are found in
at most 3 different packages (frequency 3). This characteristic makes identifiers ideal
candidates to base software provenance methods upon.

However, in some cases a given identifier will not be enough to \emph{uniquely}
identify the origin software product. Hence, to explore the practical
applicability of the proposed approach as a basis for provenance detection,
we will answer the research question:
\begin{enumerate}[\bfseries RQ1]
\stepcounter{enumi}
\item \label{rq:ii} \rqtwotext
\end{enumerate}

Based on the answer to this question we can leverage the proposed approach to
efficiently reduce the search space from all the software products in the
corpus to a handful of candidates (and then potentially apply more expensive methods---such
as clone detection techniques---to find the final result).  We explore this
answer at different granularities, from individual files up to entire
packages, as well as different methods for choosing the identifiers to query
starting from the input source artifact.  We find that on average it is enough
to chose 3 \emph{non-frequent} (within the corpus) global identifiers from a given source code file to narrow
down the software origin to no more than 3 products with 89\% probability.

As our last contribution we validate the practical usefulness of the proposed
approach by using it to determine the origin of Python packages shipped by the
Debian GNU/Linux distribution. By independent means (not based on identifiers)
we establish a ground truth correspondence between 2181 packages included in
the Debian Buster release and the \pypi packages they originate from. We then
first use the proposed approach to narrow down candidate origins for each
Debian package, and then rank candidates by pertinence using
SourceRank~\cite{saini2020popmetrics}. We find that this approach returns the
correct origin as the first candidate with a recall of 0.7 and precision of 0.8 when using 3 identifiers from only one file, and
a recall of 0.8 and precision of 0.9 when using 3 identifiers from 3 different files.
Furthermore, by repeating the search at most 5
times, and only inspecting the top result in each search, the recall improves to 0.9 with a precision of 0.77.

\paragraph{Paper structure.}

We review related work in \Cref{sec:related}. \Cref{sec:approach} details the
conceptual model of software provenance that underpins the approach proposed in
this paper. The main body of empirical work is presented in
\Cref{sec:rq1,sec:rq2,sec:validation}: \Cref{sec:rq1} establishes the
uniqueness of identifiers in the studied \pypi corpus (RQ1), \Cref{sec:rq2}
measures how many identifiers are needed to narrow down the set of candidate
product origins (RQ2), and \Cref{sec:validation} validates the approach by
identifying the \pypi origin of Debian Python packages. We discuss obtained
empirical results in \Cref{sec:discussion}, including threats to their
validity. We conclude in \Cref{sec:conclusion}, where we also suggest
directions for future work.

 \section{Related work}
\label{sec:related}

Several bodies of work in the literature use identifier-based
approaches to determine software provenance. In
this section we compare and contrast them with the approach proposed in this paper.

\subsection{Software provenance}

The term \emph{provenance} denotes a body of evidence used to establish what is
the origin and the history of a development of an artifact of any kind. In the
specific context of software development, stakeholders often wish to know how a
(software) artifact came to occur, where it originated from, its evolution,
and where it moves to over time~\cite{godfrey2015understanding,rousseau2020provenance}.
The exponential increase
of available free and open source software (FOSS) has led to the prevalence of
code reuse which in turn has made software provenance an increasingly relevant
concern in software development. However, provenance recovery has received
relatively little attention in software engineering research, with few
exceptions that we discuss below.

In its most basic definition, the provenance of a software artifact is the location, within
a reference corpus, where the artifact can be found. Godfrey~\cite{godfrey2015understanding} enumerates three
common challenges that have to be faced to fully address this problem:
definition of the scope and type of artifacts one wants to track, the gap
between identified provenance and ground truth, and the need of scaling up
provenance identification algorithms to large data sets.  With respect to this
checklist, in the present study we identify the provenance of (Python) source code
entities at various granularities (package, version, file),
sampling identifiers (class and function names) contained in the entity
itself and matching them against a large-scale identifier-to-entity corpus.
Our base lookup runtime complexity is logarithmic ($O(log(n))$---using standard database indexing
technology---allowing to scale to large
corpora. Also, to
the best of our knowledge, this is the first study of software provenance
recovery that uses a small subset of identifiers
extracted from the software under audit.

Our approach is a practical instantiation of the Bertillonage framework
introduced by Davies et al.~\cite{davies2011software,davies2013software} and
Godfrey~\cite{godfrey2015understanding}. The principle of Bertillonage is using
computationally inexpensive techniques to narrow down the provenance search
space and then applying more expensive approaches like manual determination
or clone detection algorithms. With the present work we establish
empirically that identifier-based search is a viable approach, both in terms of
correctness and efficiency, for the first part (narrowing down) of
Bertillonage, without restricting the design space for the second part.

For software artifacts implemented in Java, Davies et
al.~\cite{davies2011software,davies2013software} also adopted the Bertillonage
framework to identify the origin of \texttt{.jar} bundles within a large Java
source code corpus, Maven2. As indexing keys they used anchored class
signatures, consisting of the class and method names in a class file. By
searching for all signatures of a given package in a signature corpus, they
recovered the set of packages with the highest similarity with the queried one.
We show in this paper that comparable results can be obtained using
significantly less information (a few identifiers only, instead of all
signatures in the package).  Moreover, our approach is more flexible as it can
match at different artifact granularities (packages, versions, files).

Other linked problems in the area of software provenance are addressed in the
literature. For example, Godfrey et al.~\cite{godfrey2005using} proposed to use
identifiers as a way to identify functions that have been either merged or
split across different versions of the same software. Di Penta et
al.~\cite{dipenta2010identifying} proposed a code-search approach, using
filenames and class names, to identify software licenses.
Ossher et al.~\cite{ossher2011file} analyzed file-level clones in open source Java
projects using several trivial methods: exact file matching, filename matching,
identifier name fingerprint matching, and directory matching.

Rousseau et al.~\cite{rousseau2020provenance} introduced a compact storage
model that allows to capture the provenance information of source code files in
commits, and commits in software repositories, at the scale of the Software
Heritage archive~\cite{dicosmo2017swh,pietri2019swh}, which is the largest public archive of
software source code ($\approx$10 billion files). Their approach is at much
larger scale than what we experimented with for this paper, but the supported
granularity is different.
Also, identifier-based search is robust against file changes (as long as
identifiers remain unmodified), whereas any file change makes files no longer
recognizable within Software Heritage due to the use of strong cryptographic
hashes as artifact identifiers.

Data model, ontologies, and document standards for capturing provenance in a
broader context than software (but with applications \emph{to} software
artifacts) have also been
developed~\cite{moreau2011open,missier2013d,butt2020provone+,missier2013w3c},
together with accompanying tools~\cite{dang2008code}, and
techniques~\cite{miles2011prime,wendel2010provenance}. For software,
Bose~\cite{bose2019blinker} recently proposed a blockchain-enabled framework
based on a standard model to manage provenance data at granularities ranging
from releases to individual source files.

\subsection{Code clone detection}

In the architectural view advocated by this paper, code clone detection is
useful to \emph{precisely} pinpoint the origin of an artifact, \emph{after} a
small set of candidates has been obtaining using identifier-based narrowing
down. Since applicable code clone detection techniques abound, rather
than detailing all possibilities we refer the reader to consolidated surveys in
the field~\cite{gupta2018survey,gautam2016various,sheneamer2016survey,roy2007survey}. 

The most relevant code clone detection approaches for our work are token-based
approaches, since we also rely on source code lexical tokens (identifiers). In
these approaches source code is first tokenized, then tokens are scanned
to identify code clones~\cite{kapdan2014structural}, e.g., by relying on
sub-sequence similarity. One of the major challenges of using 
these clone detection techniques for provenance identification is that they are
prone to false positives.

CCFinder~\cite{kamiya2002ccfinder} is a multilingual token-based code clone
detection system. The lexical analyzer processes a source file producing a
stream of token, in which all identifiers are normalized to a special
token. Then, a suffix tree matching algorithm is used to find similar token
sub-sequences. CP-Miner~\cite{li2006cp} is CCFinder successor and searches for
copy-pasted code blocks and copy-paste-related bugs. To that end CP-Miner uses
a threshold to detect code similarity as a percentage of unchanged identifiers.
Yang et al.~\cite{yuan2012boreas} created Boreas, an accurate and scalable
token-based code clone detection tool, which introduced a metric-based system
to capture identifiable characteristics of program segments.

Gabel et al.~\cite{gabel2010study} conducted one of the first studies on source
code uniqueness. They used a token-based approach for finding cloned fragments
within a corpus. They defined the uniqueness of a unit of source code within the
corpus as the degree to which each project can be ``assembled'' solely from
portions of the corpus. They asked research questions similar to ours
including: ``at which granularity is software unique?'' and ``at a
given granularity, how unique is software?''. In their study, the
possible granularities are defined in terms of the length of considered token
sequences rather than logical units (package, version, file, snippet) as we do. Perez et
al.~\cite{perez2019cross} proposed tree-based machine learning approach to
detect cross-language clones, preserving identifier names in their abstract
syntax trees.

A particularly relevant approach (and accompanying tool) for detecting code
clones is SourcererCC~\cite{sajnani2016sourcerercc}, which was introduced to
scale code clone detection to large code repositories---25\,K projects for a
total of 250\,MLOC in the original experimental evaluation. On the one hand, SourcererCC follows
an approach similar to ours, in particular it relies on an external code index
based on tokens extracted from code, but it addresses a different (and more
complex) problem than ours: identifying all code clone \emph{pairs}, in the
code base under analysis, e.g., to inform and support large-scale code
refactoring. On the other hand we consider our code index to be a trusted
knowledge base or preexisting open source code and use it to find where (if at
all) code in the code base under audit comes from. At the code clone detection
problem SourcererCC outperforms, in terms of scalability, all competitors at
the time, losing in accuracy only to NiCad~\cite{cordy2011nicad}. We did not
conduct a comparative benchmark to any of those tools, because we solve a
different problem. We observe that in terms of ballpark lookup times we
outperform SourcererCC---our lookups are almost instantaneous and building the
index takes hours rather than days---but that is not a fair benchmark due to
the difference in the addressed problem.

Other differences w.r.t.~our approach and SourcererCC are worth noting.  First,
SourcererCC uses as tokens almost all code lexemes, including language keywords,
whereas we only considered identifiers. Given the effectiveness we notice in
our experiments it would be worth trying to use only identifiers also in the
case of SourcererCC, to further speed up clone detection. It is unclear however
how doing so would impact accuracy. Relying only on identifiers could also make
SourcererCC more robust and language-agnostic, because identifier extraction is
something that could be performed without having to fully parse (or even just
lex) source code, as developer tools like
Ctags\footnote{\url{http://ctags.sourceforge.net/}, retrieved 2022-09-22} do.

Second, SourcererCC relies on a bag-of-words (multiset) representation, whereas
we limit ourselves to the single presence/absence of identifier tokens (set),
not \emph{counting} how many of each of them are encountered. Furthermore,
SourcererCC use token frequency as a natural ordering of tokens as the basis
for its heuristic for reducing the number of clone pair candidates to
consider. Having to solve a simpler problem we did not need to resort to
multiset representations to achieve good accuracy results. Exploring whether
they could become even better using a bag-of-words representation remains to be
explored as future work.

SourcererCC stands out w.r.t.~most of its competitors due to its
ability to handle Type-3 clones, where cloned fragments might have been
modified upon reuse. By construction our approach deals well with detecting the
provenance of code reused and subjected to Type-3 changes, as long as the
  modifications do not affect function and class identifiers (this is
generally the case, due to the fact that those identifiers constitute APIs/ABIs
which could induce breakages on unrelated software components, and renaming identifiers
in a system is an expensive and potentially buggy operation);
there are, however,
exceptions (e.g., in the case of plagiarism where malicious actors actively try
to avoid provenance/clone detection).

Finally, we would like to stress that SourcerCC (and other clone detection methods) and our approach are complementary.
Once the corpus has been created, our approach can be used to find a small set of potential candidates (true origin or
copies of the subject system being analyzed) in a matter of seconds; at this point, a clone detector can be used between the potential candidates
and the subject to properly identify the origin of the subject, thus reducing the computational requirements of the
clone detection analysis.

\subsection{Identifier names}

Finally, we review research findings on identifier names, independently from
their use in connection with software provenance. Although not strictly
relevant for our use case, this body of work sheds light on certain
characteristics of identifiers, such as their distinctiveness, that can inform
the design of identifier-based solutions for software provenance.

Deissenboeck and Pizka~\cite{deissenboeck2006concise} observed that identifier
tokens account for approximately 33\% of the tokens and 72\% of the characters
in the source code of Eclipse, making identifiers a quantitatively
relevant part of the informative code of source code. Caprile and Tonella
agree~\cite{caprile2000restructuring}, further claiming that identifier names
are one of the most important sources of information about program entities.
Interestingly, back in the 90's Sneed~\cite{sneed1996object} conversely found
that ``in many legacy systems, procedures and data are named arbitrarily
[\ldots] programmers often choose to name procedures after their girlfriends or
favorite sportsmen''.

A more convincing explanation for the distinctiveness of identifiers comes from
the fact that, with a very high probability, different programmers would name
the same entity differently, as Butler et al.~\cite{butler1995retrieving}:
``the probability of having two people apply the same name to an object (in
general not just in code) is between 7 and 18\%, depending on the object''.

It is also commonly believed that the quality of identifier names has high
correlation with software quality. For example, Binkley et
al.~\cite{binkley2013impact} stated that identifier names are at the core of
program comprehension, and the style of identifiers (e.g., abbreviation and
camelCase) has a tremendous impact on program understanding, quality, and
development cost. Similarly, Lawrie et al.~\cite{lawrie2007effective} showed
that both actual words and abbreviations in identifiers lead to better program
comprehension, while excessively long identifiers overload short-term memory
and negatively impact program comprehension, therefore a balance between
information content and recall ability in identifiers is required. Yet,
Hofmeister et al.~\cite{hofmeister2017shorter} found that shorter identifier
names take longer to comprehend, and that using words as identifier names helps
to improve software quality and save costs.

Deissenboeck and Pizka~\cite{deissenboeck2006concise} also proposed rules for
consistent and concise identifier naming by curating an identifier dictionary
during software development. Along similar lines, recent studies have shown how
to manage identifier naming with automatic approaches. For example, Arnaoudova
et al.~\cite{arnaoudova2014repent} conducted empirical research on the
programmer activity of renaming identifiers and developed a tool to
automatically document, detect and apply renames in source code. Similarly,
Warintarawej et al.~\cite{warintarawej2015software} proposed an approach to
automatically classify software identifiers.

Nguyen et al.~\cite{nguyen2020suggesting} proposed MNire, a machine learning
approach to check the consistency between the name of a given Java method and
its implementation. They also looked into the distinctiveness of method names,
discovering that in a selected set of high-quality Java open source projects
62.9\% of method names are unique, 35.9\% of them can be tokenized into
separate words, and 78.1\%
of obtained tokens are shared among method names. We conjecture that the
difference between the distinctiveness of Java method names and Python function
names is due to primarily the size difference of the corpora (their Java dataset
includes only 14\,K products, while our \pypi dataset includes 244\,K) since the probability of a collision of names is
significantly impacted by the size of the population.

Identifiers, and specifically \emph{public} identifiers that appear in exported
functions and classes like the ones we use in the present work, can also be
leveraged to improve code search results. Exemplar~\cite{mcmillan2012exemplar}
is a seminal approach on this, translating developer searches expressed in
natural language into API documentation searches (e.g., docstrings) and then
searching a code corpus indexed by public identifiers. The indexing approach is
analogous to what we use in this paper, but the problem solved very different:
we aim at detecting where code at hand comes from and we do not use as input
natural language queries, but directly the source code under audit instead.

In previous work by two of the authors~\cite{debsources-ese-2016}, a large open
dataset of identifiers was produced, by mining using Ctags (the same tool used
in this paper) the entire source code of historical releases of the Debian
distribution. It is a larger and more diverse dataset of the one produced in
this paper, but no distinctiveness analyses were conducted at the time.

 \section{Conceptual model}
\label{sec:approach}

We briefly describe our provenance recovery approach as: given a source code
entity whose provenance one wants to identify (the \emph{subject}), we sample
some of the global identifiers contained in the entity, match them against a \corpus (a curated collection of software artifacts),
and return a small set of entities that also contain all of the sampled
identifiers.
Common types of source code entities that we will consider in the following
are: individual source code files, entire source code products (or
``packages'', e.g., Apache Spark, which is released multiple times over time),
and product releases (e.g., Apache Spark 3.0, containing several source code
files).
We will refer to the set of source code entities returned by a provenance
recovery mechanism as the \emph{candidates}. Ideally the \subject should be in
the set of \candidates, (and this set be of size of one, ie. containing the
subject and the candidate be the same).

\subsection{Uniqueness of global identifiers}

Any provenance discovery requires a reference \corpus. 
A \corpus is a collection of software
entities that is curated (harvested, preprocessed and cleaned-up) to serve as a reference for provenance discovery.
Finding the origin of a software entity (the \subject) means
finding the entity in the \corpus from which the \subject was copied from.\footnote{Note that this copy might not have
  been done directly from the corpus; it is, however, a copy of the same entity that exists in the corpus.} 

A \corpus can be created by scanning and downloading the source code of products from any repository from which there is some confidence that
the product comes from its true creator/maintainer. For example, a \corpus can be created from version control repositories
(such as GitHub and GitLab), or from repositories of components  used for dependency management  (as long as they include source code, such
as Maven Central, NPM, \pypi, CRAN, etc). Ideally a repository should be as comprehensive as possible.

To be able to precisely assert that a \subject entity is a copy of a specific \candidate entity in the \corpus would require having
information that documents how the \subject entity was copied from the corresponding \candidate entity in the
\corpus.  Even if this was possible, there will still be instances where identical entities might have been created
independently of each other (for example, the identifier \texttt{main} is created in every single C program, sometimes
it might be a copy from another program, sometimes it will be typed in---created from scratch---by the programmer).
Thus a given \subject entity might have different potential matches (\candidates entities).

Another aspect to consider when designing the \corpus is how the location of the \candidate is reported. Software
entities exist in a hierarchical structure (recursively, an entity is composed of other entities). If there is a match
for a given \subject entity in the \corpus, this result can be reported at any level of containment. For example, if the
\corpus was built from releases of products, the \candidates' location can be reported as: the product (e.g., Apache
Spark),  a release of the component (Apache Spark 3.0) a file in a specific
release, and even the specific location in the source code file where entity is defined. If the \corpus was built from version control systems, it can be
reported as a URL of the GitHub repository, as a revision/tag in this repository and a specific file in a specific
revision.

In the context of this research, we are concerned with matching source code global identifiers defined inside an entity (e.g.,
software product, one of its releases, or one of the files in one of its releases). We use the notation $Defs(e)$ to
denote the set of identifiers defined in the entity $e$, where $e$ can be a
product $P$, a release $R$ or a file $F$. This function is computed taking into consideration the
containment relationship of the entities. For instance, the identifiers defined in a product are the union of the
identifiers in its releases:
\[ Defs(P) = \bigcup_{R_i \in P} Defs(R_i)\]
and the  identifiers in a release are the union of the identifiers of its files:
\[ Defs(R) = \bigcup_{F_i \in R} Defs(F_i)\]

A \corpus is created by defining the entity of interest (in particular, for the purpose of this research, an entity
is the set of all releases of a software product). Such entity corresponds to a \emph{document}
in the information retrieval nomenclature\cite{introIR}, thus, an identifier $id$ is in a document $e$ (entity e)  iff  $id \in Defs(e)$.
Thus, in this research, the document frequency of an identifier in the \corpus is the number of software products in the \corpus
in which the identifier is defined.

For the sake of readability, in the rest of this document, we will refer to the document frequency of an identifier as
the frequency of such identifier (within a given \corpus).

 \section{On the uniqueness of identifiers}
\label{sec:rq1}

The first research question we address is:

\begin{enumerate}[\bfseries RQ1]
\item \rqonetext
\end{enumerate}

For the purpose of provenance and origin analysis, we only consider global identifiers. That is, identifiers that
can either be referenced from other programs, or that identify entities that could potentially be copy-pasted to
other software.

We have chosen Python for our empirical experiments, and more specifically \pypi (the Python Package
Index) for several reasons.
First, Python is one of the most popular programming languages today with an active and vibrant ecosystem. Second, \pypi is
an authoritative index of Python products and as such, it can be considered the canonical directory of most Python
projects.\footnote{Projects do not reside in \pypi, but \pypi links to their actual location.}
And third, it is comprehensive: as of Sept. 2021, \pypi lists 331\,K Python projects
and 2.9\,M releases of them.

The Python language does not have access modifiers; all identifiers defined at the top-level are publicly available (this is in contrast to languages like
Java and C where the programmer can decide if an identifier is public, private or protected). We build our
corpus with global functions and global classes (and their methods) as they are the most likely types of identifiers
that are referenced (both internally within a product, and across products). For the rest of this paper, we will use the term
function  to
refer to both functions or methods.
We use identifiers that are \emph{not qualified} by the class (in the case of methods) or module they belong to.
In Python filenames correspond to names in the module hierarchy---e.g., a file named \texttt{foo.py} can be imported as \texttt{import foo}---the (path-less) name of a source code file is another identifier that can be used for provenance discovery.

Most Python products have many releases, and most of their identifiers will
appear in multiple releases. For the purpose of answering RQ1, we will focus on
identifying the \emph{product} where identifiers are defined in, rather then the
specific \emph{releases} of such products. With respect to the conceptual model of
\Cref{sec:approach}, the scope of an entity will be a product, a document is a product,
and the
document frequency of an identifier corresponds to the number of \pypi products in which
such identifier is defined (in at least one release).

\subsection{Methodology}
\label{sec:rq1method}

Our methodology for answering RQ1 consists of three main steps:
\begin{enumerate}
\item \textbf{Source code retrieval}: Crawling and downloading each release of every package in \pypi
\item \textbf{Identifier extraction}: For each package: extract all the identifiers defined in all its releases 
\item \textbf{Frequency measurement}: Measure the frequency of every identifier found, at product granularity
\end{enumerate}
We further detail each step below.

\subsubsection{Source code retrieval}

We performed the source code retrieval on August 20, 2020, using the following method:

First, we retrieved the list of products with \pypi-simple (the official \pypi
API.\footnote{\url{https://pypi.org/project/pypi-simple/}, accessed 2021-10-25}
\pypi-simple returned \num{244 084} products (same as the number displayed on
\url{https://pypi.org}) with a total of 1.83 million releases.

Next, for each product we downloaded its metadata and list of releases (using
\pypi JSON's
API\footnote{\url{https://warehouse.pypa.io/api-reference/json.html}, accessed
  2021-10-25}). Some products had been retired and had no further information
available on \pypi, other were not associated to any downloadable files; in
total \num{13 379} products did not contribute any downloaded release to our
corpus for these reasons.

\begin{table}
  \centering
  \caption{Descriptive statistics of \pypi products and their releases}
  \begin{tabular}{lr}
    \toprule
    \# products:                                    & \num{244 084}         \\
    \# releases:                                    & \num{1 831 172}       \\
    \# products with only 1 release:                & \num{69 306} (28.4\%) \\
    \# products with more than 100 releases:        & \num{1 451} (0.6\%)   \\
    \# products without Python source files:        & \num{13 739} (5.6\%)  \\
    \midrule
    \                                               & Median / Mean / Stdev \\
    \# releases per product:                        & 3 / 7.5 / 13.0        \\
    \# files per release:                           & 6 / 24.9 / 137.7      \\
    \midrule
    \# files:                                       & \num{45 239 359}      \\
    \# different base filenames:                    & \num{1 117 588}       \\
    \bottomrule
  \end{tabular}
  \label{tab:stats_corpus}
\end{table}

The next step was to download, for each of the products, their releases. We observed that most products had few
releases (the median number of releases per product was 7.5, with a standard deviation of 13). However, few products had an unusually high number of releases
(for example, \package{CCXT}, a real-time cryptocurrency trading library, had over \num{7400} releases in 3 years). For this
reason we decided to only download the 100 most recent releases of a product. Only 0.6\% of products had more than 100
releases.

To download a release we retrieved one of its source distribution archive files (\pypi \texttt{packagetype = "sdist"}). These
types of files contain the complete source code of the release. A release can be offered in several formats, all with
the same contents (such as \texttt{.zip}, \texttt{.bz2}, \texttt{.tgz}, etc.); therefore, we downloaded the first one
that was listed. If the release did not provide a source distribution, we downloaded one of the binary distribution files
(\texttt{packagetype = "bdist\_wheel"} or \texttt{"bdist\_egg"}). Being Python
an interpreted languages, ``binary'' distributions can still contain Python source
code files,
but only those source files necessary to run the software; it is likely that they do not contain certain types of
files, such as those used during testing or building.  We ignored 132 releases  that did not have source distributions nor binary distributions, but only installer files such as \texttt{.exe} and
\texttt{.rpm}. 17 products did not include any files for download. \Cref{tab:stats_corpus} summarizes 
results of this step.

The downloaded compressed releases occupied 1.6\,TBytes of disk space.

\subsubsection{Identifier extraction}

We used \verb|Universal Ctags|\footnote{\url{https://ctags.io/}, accessed 2021-10-25. 
Universal Ctags 0.0.0 (2015) derived from Exuberant Ctags 5.8.} to extract class and function
identifiers from Python source code.
We did not transform the identifiers in any way (we did not normalize
capitalization, nor split compound tokens in \texttt{CamelCase} or
\texttt{snake\_case} conventions).

For each release, we uncompressed its files into a temporary directory and ran \texttt{ctags} recursively on all files of the release
with extension \texttt{.py} (case insensitive). We discarded all identifiers except classes (type \emph{class} in \texttt{ctags} results) and
functions (types \emph{function} and \emph{member}). For filenames, we discarded their path.
For each identifier we recorded a tuple
$\langle$\emph{product~name}, \emph{identifier~name}, \emph{identifier~type}$\rangle$ in a \texttt{sqlite3}
database. Our database is 66 Gbytes (including indexes).

\begin{table}
  \centering
  \caption{Descriptive statistics for the identifier corpus. All numbers correspond to unique names.}
  \begin{tabular}{lr}
    \toprule
    \# class names: & \num{2 665 927} \\
    \# function names: & \num{8 598 979} \\
    \# filenames: &  \num{1117588}\\
    \                               & Median / Mean / Stdev \\
    \# function names per product: & 21 / 92.0 / 471.8 \\
    \# function names per release: & 40 / 155.0 / 629.5 \\
    \# class names per product: & 6 / 29.1 / 184.3 \\
    \# class names per release: & 10 / 46.3 / 210.6 \\
    \# filenames per product: & 4 / 12.81 /65.6 \\
\bottomrule
  \end{tabular}
  
  \label{tab:stats_corpus_idents}
\end{table}

We identified that \num{13739} products (5.6\%) did not include any Python files, and therefore did not
contribute any identifiers to our corpus. Many of these products were written in languages other than Python.

Table~\ref{tab:stats_corpus_idents} summarizes the main statistics of the extracted identifiers.  We found 2.6 million
different class names, 8.6 million function names, and 1.1 million different filenames in the \pypi corpus.  As it can be
seen, most products have very few filenames (median 4 different filenames), and define a relatively small number of
identifiers (median: 6 different class names and 21 different functions names).

Using a SQLite database, for each identifier (function, class or filename), we computed its distinctiveness (the
number of different products where it was found).

\subsection{Results}

\subsubsection{Frequency measurement}

\begin{table}
  \setlength{\tabcolsep}{5pt}
  \centering
  \caption{Distribution of frequency of identifiers at product level}
  \begin{tabular}{l|rrr|rrr|rrr}
    \toprule
             & \multicolumn{3}{c}{Class names} & \multicolumn{3}{|c|}{Function names} & \multicolumn{3}{c}{Filenames}            \\
    \cmidrule(r){2-4}\cmidrule(l){5-7}\cmidrule(l){8-10}
    frequency     & \#                           & Prop & Cum  & \#     & Prop & Cum & \#  & Prop & Cum\\
             &  Idents                         & (\%) &(\%) &  Idents    & (\%) &  (\%) & Filenames & (\%)& (\%)\\
    \midrule
    1        & \num{2 020 027}                       & 75.8 & 75.8     & \num{6 595 770} & 76.7 & 76.7      & \num{886 236} & 79.3 & 79.3  \\
    2        & \num{343 195}                         & 12.8 & 88.6     & \num{1 048 985} & 12.2 & 88.9      & \num{122 929} & 11.0 & 90.3  \\
    3        & \num{117 016}                         & 4.4  & 93.0     & \num{335 572}   & 3.9  & 92.8      & \num{43 064}  & 3.9  & 94.2  \\
    4        & \num{52 326}                          & 2.0  & 95.0     & \num{164 056}   & 1.9  & 94.7      & \num{17 234}  & 1.5  & 95.7  \\
    5        & \num{30 224}                          & 1.1  & 96.1     & \num{101 079}   & 1.2  & 95.9      & \num{10 038}  & 0.9  & 96.6  \\
    6        & \num{20 528}                          & 0.8  & 96.9     & \num{60 735}    & 0.7  & 96.6      & \num{6 458}   & 0.6  & 97.2  \\
    7        & \num{12 829}                          & 0.5  & 97.4     & \num{44 863}    & 0.5  & 97.1      & \num{4 638}   & 0.4  & 97.6  \\
    8        & \num{9 786}                           & 0.3  & 97.7     & \num{35 849}    & 0.4  & 97.5      & \num{3 273}   & 0.3  & 97.9  \\
    9        & \num{8 622}                           & 0.4  & 98.1     & \num{28 282}    & 0.2  & 97.9      & \num{3 004}   & 0.3  & 98.1  \\
    10       & \num{6 623}                           & 0.3  & 98.3     & \num{22 702}    & 0.2  & 98.1      & \num{2 886}   & 0.3  & 98.4  \\
    11-100   & \num{42 241}                          & 1.6  & 99.9     & \num{150 015}   & 1.7  & 99.9      & \num{15 885}  & 1.4  & 99.8 \\
    101-1000 & \num{2 385}                           & 0.1  & 100      & \num{10 308}    & 0.1  & 100       & \num{1 795}   & 0.2  & 100   \\
    1001-    & \num{125}                             & 0.0  & 100      & \num{763}       & 0.0  & 100       & \num{148}     & 0.0  & 100   \\
    \bottomrule
  \end{tabular}
  \label{tab:dist}
\end{table}

\begin{table}
  \centering
  \caption{Distribution of frequency at product level of source code identifiers combined}
  \begin{tabular}{lrrr}
    \toprule
    & \multicolumn{3}{c}{Source code identifiers}\\
    & \multicolumn{3}{c}{classes and methods}\\
    & \multicolumn{3}{c}{combined}\\
    frequency & \# Idents & (\%) & Cum (\%) \\
    \midrule
    1           &           \num{8561214}     &           76.4        &           76.4      \\
    2           &           \num{1385535}     &           12.4        &           88.8      \\
    3           &           \num{451270}      &           4.0         &           92.8      \\
    4           &           \num{215288}      &           1.9         &           94.8      \\
    5           &           \num{130731}      &           1.2         &           95.9      \\
    6           &           \num{81036}       &           0.7         &           96.7      \\
    7           &           \num{57359}       &           0.5         &           97.2      \\
    8           &           \num{45546}       &           0.4         &           97.6      \\
    9           &           \num{36750}       &           0.3         &           97.9      \\
    10          &           \num{221186}      &           0.3         &           98.2      \\
    10--100     &           \num{191844}      &           1.7         &           99.9      \\
    101-1000    &           \num{12778}       &           0.1         &           100.0     \\
    1001 -      &           \num{902}         &           0.01        &           100.0     \\
    \bottomrule
  \end{tabular}
  \label{tab:distboth}
\end{table}

Table~\ref{tab:dist} summarizes the results for each type of identifier. 75.8\% of class names, 76.7\%
of function names, and 79.3\% of filenames exist in only one product. 
Class and function identifiers have almost identical distributions: in both cases, 93\% of
identifiers have a frequency of at most 3. The distribution for filenames  is very similar but with two notable
differences: it has slightly more unique filenames, and almost twice as many very common filenames (frequency $>$100) than both class and function names.

\subsubsection{Identifiers used for both: classes and function names}

The intersection between the names of classes and names of functions is very small: only
\num{65 982} identifiers (out of 11.2 million identifiers, 0.59\%) are used for both classes and functions. In other
words, in 99.41\% cases,
the name of the identifier is sufficient to know if it is a class name or a function name.
\Cref{tab:distboth} shows the frequency of the combined source code identifiers. While comparing this
table to \Cref{tab:dist}, note how
the distributions of class names, functions names, and the combined identifiers (both class names and function names)
are virtually identical. 

These results allow to conclude that we do not need to use the type of identifier to determine the product where it is
declared. Thus, for the rest of the paper we will use the term identifier to refer to both a function or
class name identifier.

Filenames have a significantly larger intersection with other types of identifiers: 12\% of filenames are also a class
name, and 17\% a function name.

\subsubsection{Probability of sampling a unique identifier or filename}

The results of the previous section indicate that more than 75\% of identifiers and filenames are unique. However, some
identifiers are very common. Frequent identifiers might be much more common than all the unique identifiers
combined. For this reason, it is important to calculate the proportion of instances corresponding to each
frequency.  This is similar to conducting the following experiment: if we had a set of all instances of all
identifiers defined in the corpus (name of product, identifier), and were to randomly choose an identifier from
this set, how many products would have the same identifier? (i.e. what would be the frequency of this identifier).  We
repeat this experiment for filenames too.

In the \pypi corpus there are 26.8 Million instances of identifiers, but only 11.2 M different ones (e.g., the
identifier \emph{main} was declared in \num{53 413} products). For filenames there are 2.9 M instances, 1.1 M different ones.
\Cref{tab:dist_inst} shows the proportion of total instances of identifiers that
have a given frequency. For instance, at frequency 2 (i.e., identifiers that occur in two different
products), there are \num{1 390 971} different identifiers that occur in \num{2 781 942} products, which correspond to
10.4\% of all identifier instances in the corpus.

Even though the frequency of files and identifiers is not that different, those differences compound. As shown in
table~\ref{tab:dist_inst}
only 32.1\% of identifier instances are unique, and 
50\% of identifiers instances have a frequency of 4 or less.
If we were to randomly choose an identifier within a randomly
chosen product, the probability that this identifier is unique is 32.1\%;
and in 50\% of the cases, its frequency would be 4 or less. 
For filenames, the probability of randomly choosing a unique one is
30\%; and in 50\% of the cases, it would have a frequency of 8 or less.
The most common identifiers (frequency $>1000$), even though only 888, account for 10.9\% of all instances.
In the case of files, 148 filenames account for 18.4\% of all instances.

\begin{table}
  \centering
  \caption{Distribution of distinctiveness of instances of identifiers and filenames at product level. For example, identifiers of
    distinctiveness equal to 4 correspond to 3.2\% of all identifier instances, and 50.8\% of instances have a 
  distinctiveness  of at most 4.}
  \begin{tabular}{l|r|rrr|r|rrr}
    \toprule
    & \multicolumn{4}{c|}{Source code identifiers} & \multicolumn{4}{c}{Filenames}  \\
    & \multicolumn{4}{c|}{(classes and methods)}\\
    \toprule
            & \# Ids & \multicolumn{3}{c|}{Instances} & \# Ids & \multicolumn{3}{c}{Instances} \\
    Frequency     &        & \#     & Prop  & Cum &        & \#     & Prop  & Cum\\
             &        &        & (\%) &  (\%) &        &        & (\%) &  (\%)\\
    \midrule
    1        & \num{8 600 252} & \num{8 600 252} & 32.1 & 32.1     & \num{886 236} &  \num{886 236} & 30.0 & 30.0\\
    2        & \num{1 390 971} & \num{2 781 942} & 10.4 & 42.5     & \num{122 929} &  \num{245 858} & 8.3  & 38.3\\
    3        & \num{452 285}   & \num{1 356 855} & 5.1 & 47.6      & \num{43 064}  &  \num{129 192} & 4.4  & 42.7\\
    4        & \num{216 279}   & \num{865 116}   & 3.2 & 50.8      & \num{17 234}  &  \num{68 936}  & 2.3  & 45.1\\
    5        & \num{131 248}   & \num{656 240}   & 2.5 & 53.3      & \num{10 038}  &  \num{50 190}  & 1.7  & 46.8\\
    6        & \num{81 232}    & \num{487 392}   & 1.8 & 55.1      & \num{6 458}   &  \num{38 748}  & 1.3  & 48.1\\
    7        & \num{57 686}    & \num{403 802}   & 1.5 & 56.6      & \num{4 638}   &  \num{32 466}  & 1.1  & 49.2\\
    8        & \num{45 620}    & \num{364 960}   & 1.4 & 58.0      & \num{3 273}   &  \num{26 184}  & 0.9  & 50.1\\
    9        & \num{36 885}    & \num{331 965}   & 1.2 & 59.2      & \num{3 004}   &  \num{27 036}  & 0.9  & 51.0\\
    10       & \num{29 308}    & \num{293 080}   & 1.1 & 60.3      & \num{2 886}   &  \num{28 860}  & 1.0  & 51.9\\
    11-100   & \num{192 252}   & \num{4 682 191} & 17.5 & 77.8     & \num{15 885}  &  \num{421 320} & 14.3 & 66.2\\
    101-1000 & \num{12 695}    & \num{3 030 540} & 11.3 & 89.1     & \num{1 795}   &  \num{453 591} & 15.4 & 81.6\\
    1001-    & \num{888}       & \num{2 913 499} & 10.9 & 100.0    & \num{148}     &  \num{543 811} & 18.4 & 100.0\\
    \bottomrule                                                    
  \end{tabular}
  \label{tab:dist_inst}
\end{table}

\subsubsection{Frequent identifiers}
\label{sec:freqidents}

\begin{table*}
  \setlength{\tabcolsep}{3pt}
  \centering
  \caption{The 10 most frequent identifier names in functions and classes. \#Prs is the number of products where that
    identifier is defined (its distinctiveness). }
  \label{tab:freq_ids}
  \begin{tabular}{lrr|lrr|lrr}
    \toprule
    \multicolumn{3}{c}{Classes} & \multicolumn{3}{c}{Functions} & \multicolumn{3}{c}{Filenames}                                               \\
    \cmidrule(r){1-3}\cmidrule(l){4-6} \cmidrule(l){7-9}
            & \multicolumn{2}{c|}{Products} & & \multicolumn{2}{c|}{Products} & Name & \multicolumn{2}{c}{Products} \\
Name                     & Freq       & \% &            & Freq       & \%  &  &  Freq & \%                            \\
    \midrule
Meta                 & \num{9 771}   & 4.0  & \_\_init\_\_    & \num{159 528} & 65.3 & \_\_init\_\_ & \num{60482} & 26.3 \\
Command              & \num{6 246}   & 2.6  & main            & \num{53 368}  & 21.9 & setup        & \num{54462} & 23.6 \\
Config               & \num{6 113}   & 2.5  & run             & \num{46 158}  & 18.9 & utils        & \num{31149} & 13.5 \\
Migration            & \num{6 062}   & 2.5  & \_\_str\_\_     & \num{44 664}  & 18.3 & cli          & \num{15600} & 6.8  \\
Client               & \num{5 520}   & 2.3  & \_\_repr\_\_    & \num{41 796}  & 17.1 & exceptions   & \num{14888} & 6.5  \\
PostInstallCommand   & \num{5 100}   & 2.1  & get             & \num{33 024}  & 13.5 & models       & \num{13816} & 6.0  \\
PostDevelopCommand   & \num{4 737}   & 1.9  & \_\_call\_\_    & \num{31 053}  & 12.7 & base         & \num{13048} & 5.7  \\
EggInfoCommand       & \num{4 598}   & 1.9  & setUp           & \num{30 205}  & 12.3 & config       & \num{9925 } & 4.3  \\
User                 & \num{4 291}   & 1.8  & read            & \num{25 981}  & 10.6 & main         & \num{9658 } & 4.2  \\
Error                & \num{4 094}   & 1.7  & \_\_getitem\_\_ & \num{25 714}  & 10.5 & util         & \num{9317 } & 4.0  \\
\bottomrule
  \end{tabular}
\end{table*}

The last row of \Cref{tab:dist_inst} shows that there are only 888 identifiers (0.01\%) with frequency larger
than \num{1000}, but they are very frequent and correspond to 10.9\% of all instances. For filenames, there are 148 files
that account for 18.4\% of all instances.
This result suggests the possibility of creating a small list of most frequent identifiers that can be excluded from any
method to determine their origin (similar to stop words in natural language processing).

The top 10 of each identifier type are presented in \Cref{tab:freq_ids}. As it can be seen, most of class identifiers are
installation- and configuration-related. They are required to be defined by a product that uses Python distribution
mechanism (used by \package{setuptools}). Regarding the most common functions, many are prefixed and suffixed with \_\_
(e.g. \code{\_\_getitem\_\_}). Python uses such identifiers---called ``dunders'' or ``magic methods''---to modify its run-time behaviour (e.g., the method \code{\_\_getitem\_\_}
of a class is used to redefine the behaviour of [\ ] indexing). Regarding files, they also correspond to common naming conventions
(e.g., \code{setup}, \code{main}, \code{config}, and \code{exceptions}).

\begin{hassanbox}{\columnwidth}[4ex]
  To summarize: in \pypi 75.8\% of class identifiers, 76.7\% of function identifiers and 79.3\% of filenames are defined in only one
  product. The set of identifiers for functions and the set of identifiers for classes are almost mutually exclusive
  (0.59\% of identifiers are used for both).  While 95\% of identifiers (and 96\% of filenames) 
  are defined in at most 4 products, these identifiers correspond to only 50.8\% of all instances
  of identifiers (45.1\% for filenames).
\end{hassanbox}

 \section{Using identifiers to determine software provenance}
\label{sec:rq2}

We have determined that most of the identifiers in \pypi are very distinct at product level. This fact makes identifiers
a promising building block for a lightweight approach to determine the provenance of a file or set of
files within a software corpus (like \pypi). In order to move from this potential to a working approach, however, we need
to turn this intuition into a practical heuristic.

Consider the following scenario: we have a set of files (we will refer to them as the subject files) from a release of
an unknown product, and we would like to determine which product they belong to. This problem can be answered with many
different methods. For example, one can create a database of hashes of every file of every release of every product in
the database. If the subject files have not been modified, their hashes can be quickly compared to this database.
However, if the files have been modified (even by a single byte), this method would not work. Alternatively, we can use
\emph{diff} tools or clone detection tools; these methods will be more expensive methods.

Answering RQ1 we established that the identifiers defined in products in the \pypi corpus are relatively
unique. We can use this information to create a ``fingerprint'' that can be used to reduce the search space of potential
products from which the subject files might have originated. If this set is small, other time-consuming methods can then be
used to match the subject files to the products in this set.

This \textbf{\fingerp} is a set of $N$ globally declared identifiers in the subject files; if we randomly extract $N$
globally declared identifiers from subject files, we expect that very few projects (potentially only one) will have declared
\emph{all} these identifiers. Thus, we can quantify the effectiveness of this method by answering RQ2: 

\begin{enumerate}[\bfseries RQ1]
\stepcounter{enumi}
\item \rqtwotext
\end{enumerate}

\subsection{Methodology}

For this experiment we used the \pypi corpus described in \Cref{sec:rq1}.  The main parameter to this experiment is the
number of distinct identifiers extracted from the subject files (the size of the \fingerp). We use $N$ to refer to this
parameter. We want to find the minimum number of identifiers needed. As we observed in the previous section, most
identifiers are unique, but there are more instances of non-unique identifiers than instances of unique identifiers.  If
we randomly choose one identifier from a package, we have a 32\% chance of finding a unique identifier (and 50\% that it
has a frequency of at most 4 packages). Thus, we are
likely to need to sample more than one identifier from the subject package to increase this probability. Our goal is to
identify the ideal number of identifiers we need. Thus, we repeat the experiment for $N\in [1,\ldots,5]$ and compare the
results.  The methodology for each value of $N$ consists of 3 parts: sampling, fingerprinting, and matching.

\paragraph{Sampling.}

We start by randomly selecting a sample of 1000 different product \emph{releases} from the corpus, making sure they all
belong to difference products. From each chosen release we extract a \fingerp (discussed below) composed of $N$ distinct
identifiers. Note that it is not always possible to do so (e.g., in extreme cases a release might contain less than $N$
distinct identifiers); in those cases the release will be discarded and another one will be chosen at random until success.

\paragraph{Fingerprinting.}

We tested two different strategies to create the \fingerp that depend on how many different files the identifiers are
sampled from. Intuitively, extracting the identifiers from a single file will result in more candidates than extracting
them from different files. Nonetheless, we are interesting to quantify how much better one method is that the other.

\begin{description}

\item \textbf{Single-file strategy:} this method requires that all $N$ fingerprint
  identifiers come from a single file. To that end we randomly select one
  source code file from the input release and then randomly select from it $N$
  distinct identifiers. If the file does not contain enough distinct
  identifiers we backtrack and pick another file. If no files allow to satisfy
  the criteria the release is discarded and another one chosen at random until
  success.

\item \textbf{Disjoint-files strategy:} this method requires that the $N$ fingerprint
  identifiers be different and come from $N$ different \emph{files} in the
  input release (i.e., one per file). To achieve this we randomly sample $N$
  files without replacement and randomly select from each file one identifier
  that has not been selected yet. In case a file does not contain any (new)
  identifier, we backtrack and select another file at random until success.  As
  before, if the release does not have enough files or enough identifiers to
  create the \fingerp, it is discarded and another one is chosen at random until
  success.

\end{description}

In term of fingerprint sizes we make $N$ vary in the $n\in [1,\ldots,5]$ closed
interval; while even the maximum size allowed by this choice (5 identifiers)
appears small, our results show that it is more than enough in practice in most
cases.

Note that there is no need to redo the sampling step for each possible value of
$N$. One can perform a single sampling with $n=5$ and then use fingerprint
\emph{prefixes} of length $n\in [1,\ldots,5]$ for the matching phase.
Similarly, in order to speed up sampling, we have excluded all releases that do
not have either enough identifiers or enough files as a whole to potentially
permit sampling 5 identifiers. This introduced some bias in the experiment, but
we believe that a useful real-world Python product will contain at least 5
identifiers.

\paragraph{Matching.}

For matching input releases to the corpus, the approach is straightforward: given a fingerprint of $N$ identifiers find
the set of all \pypi products that contain \emph{all} identifiers in the fingerprint. These products constitute the
candidate products from where the input release originate from and it is very efficient to determine. We will analyze
and discuss the number of products returned using different fingerprint sizes and sampling techniques.

\paragraph{Dealing with stop words.}

As discussed in \Cref{sec:freqidents}, some identifiers are very common and correspond to a large proportion of
the occurrences (e.g., the top 888 most frequent identifiers account for 10.9\% of all occurrences, and the top 148 filenames
for 18.4\%). Very common identifiers will not be very useful for provenance discovery.
Examples of such
identifiers for Python are \texttt{\_\_init\_\_} (the prescribed name of class constructor methods) or \texttt{main} (a
common name for the entry point function in an executable).  As it is common practice in language-based searches, we
establish a \bl: a list of the most frequent stop ``words'' (identifiers in this case) that are excluded from sampling, and therefore cannot
appear in fingerprints. An
important question is how large should the \bl be.  Intuitively, the larger the list the better (e.g., if our \bl is
composed of any identifier that exists in two or more products, then any identifier not in the \bl would uniquely
identify a product in \pypi---assuming the identifier exists in the corpus). However, a large list has two disadvantages: a) the
larger the list the more expensive it is to use it; and, more importantly b) if the list is too big,
it might not be possible to extract sufficient identifiers from the subject file to create the \fingerp.

\begin{figure}
  \centering
  \includegraphics[width=0.95\textwidth]{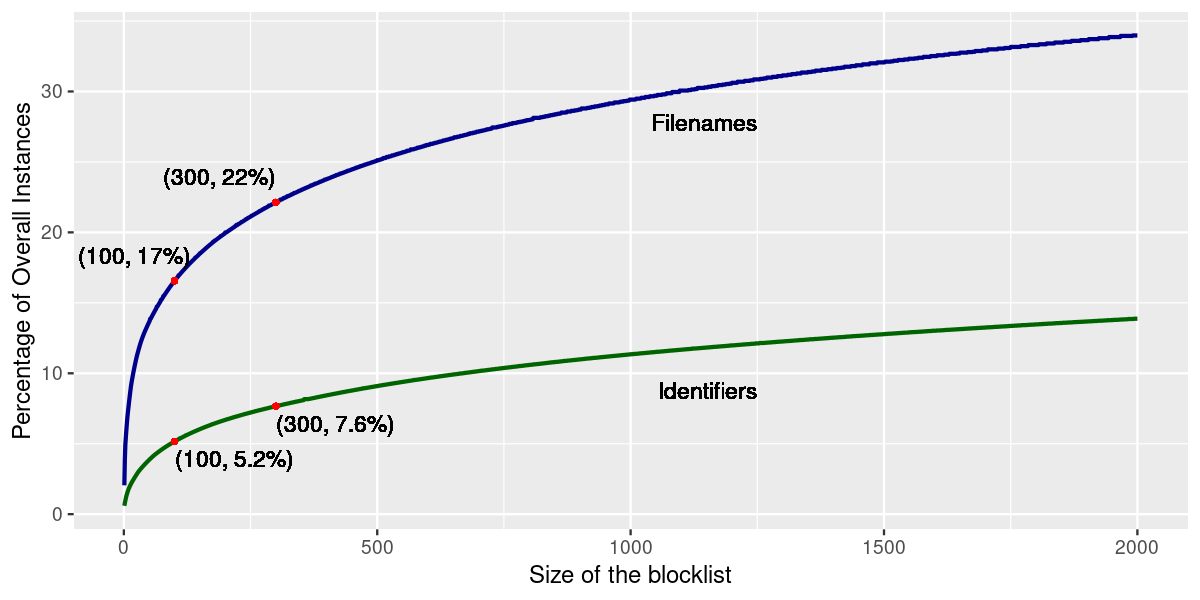}
  \caption{Impact of the size of the blocklist in the proportion of occurrences that would be removed from the corpus.}
  \label{fig:blocklist}
\end{figure}

\Cref{fig:blocklist} shows the impact of the size blocklist with respect to the proportion of identifiers removed
from the corpus (for example, the identifiers in a blocklist of size 100 account for 5.2\% of the occurrences of
identifiers; for filenames, a size of 100 corresponds to 17\% of occurrences).  We empirically verified the impact of the
blocklist by running the following experiment: we randomly sampled \num{5000} products from the corpus, using the single
file strategy, with a fingerprint size of 3---that is, an eligible product must have at least one file with at least 3
different identifiers not in the blocklist being tested; we then counted the number of products that match such fingerprint.
As for the blocklist itself, we experimented with different
sizes of it, each time corresponding to the top-$N$ identifiers in the entire corpus. We made $N$ (the size of the
blocklist) vary from 0 to \num{10 000}. 

\begin{figure}
  \centering
  \includegraphics[width=0.7\textwidth]{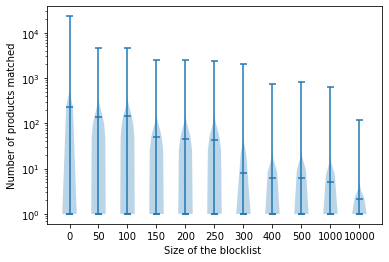}
  \caption{For each size of the blocklist, 5000 random fingerprints were matched. Thus, each column represents a
    violin-plot of the number of products match. We chose a size of 300 because it shows the last significant drop in
    the median number of matched products.}
  \label{fig:benefit_blist}
\end{figure}

Results are shown in \Cref{fig:benefit_blist}. As expected, increasing the size of the blocklist reduces the
number of candidates. There is a significant drop at 300, which is also observable in
\Cref{fig:blocklist}. 
For these reasons we decided to use a blocklist of 300 for the following experiments.  It is worth noting that by
removing the top-300 identifiers, only 0.27\% of all unique identifiers are removed; yet, this list removes 7.6\% of all identifier occurrences in the \pypi corpus. For
filenames, 300 correspond to 0.27\% of distinct filenames, and removes 22\% of all filename occurrences.

Note that a blocklist can be implemented by keeping the frequency of each identifier (or its inverse document frequency \emph{idf}) in the corpus database.
An identifier will be part of the blocklist if its frequency is above a certain threshold. In this corpus,
that threshold is at least \num{2230} (\emph{idf} 2.039)  for identifiers and \num{551} for files (\emph{idf} 2.645).

To summarize, we performed the experiments described using:
\begin{itemize}
\item two sampling strategies: single-file v. disjoint-files;
\item fingerprint size varying from 1 to 5;
\item a \bl size composed of the 300 most popular identifiers.
\end{itemize}

\subsection{Results}

Our goal is to understand how the size of the \fingerp impacts the number of products matched in the corpus.
In terms of information retrieval metrics, this experiment will always have a recall of 100\% since the project
from where the identifiers are extracted is always in the result. The precision would be the inverse of the number of
matched projects.

\begin{figure}
  \centering
  \includegraphics[width=0.8\textwidth]{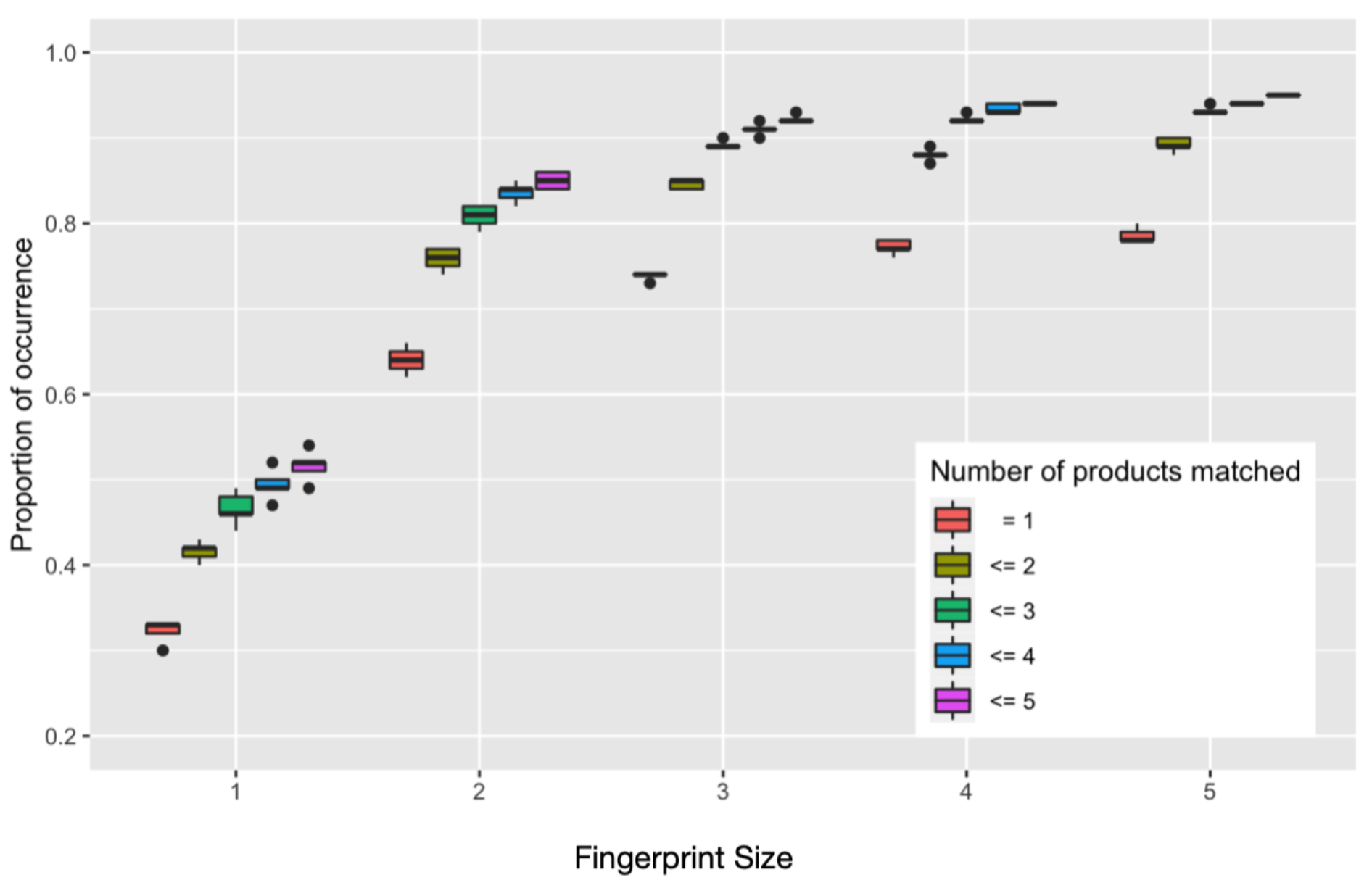}
  \caption{Distribution of the number of candidates: single file strategy
  }
  \label{fig:rq2_b}
\end{figure}

\begin{figure}
  \centering
  \includegraphics[width=0.8\textwidth]{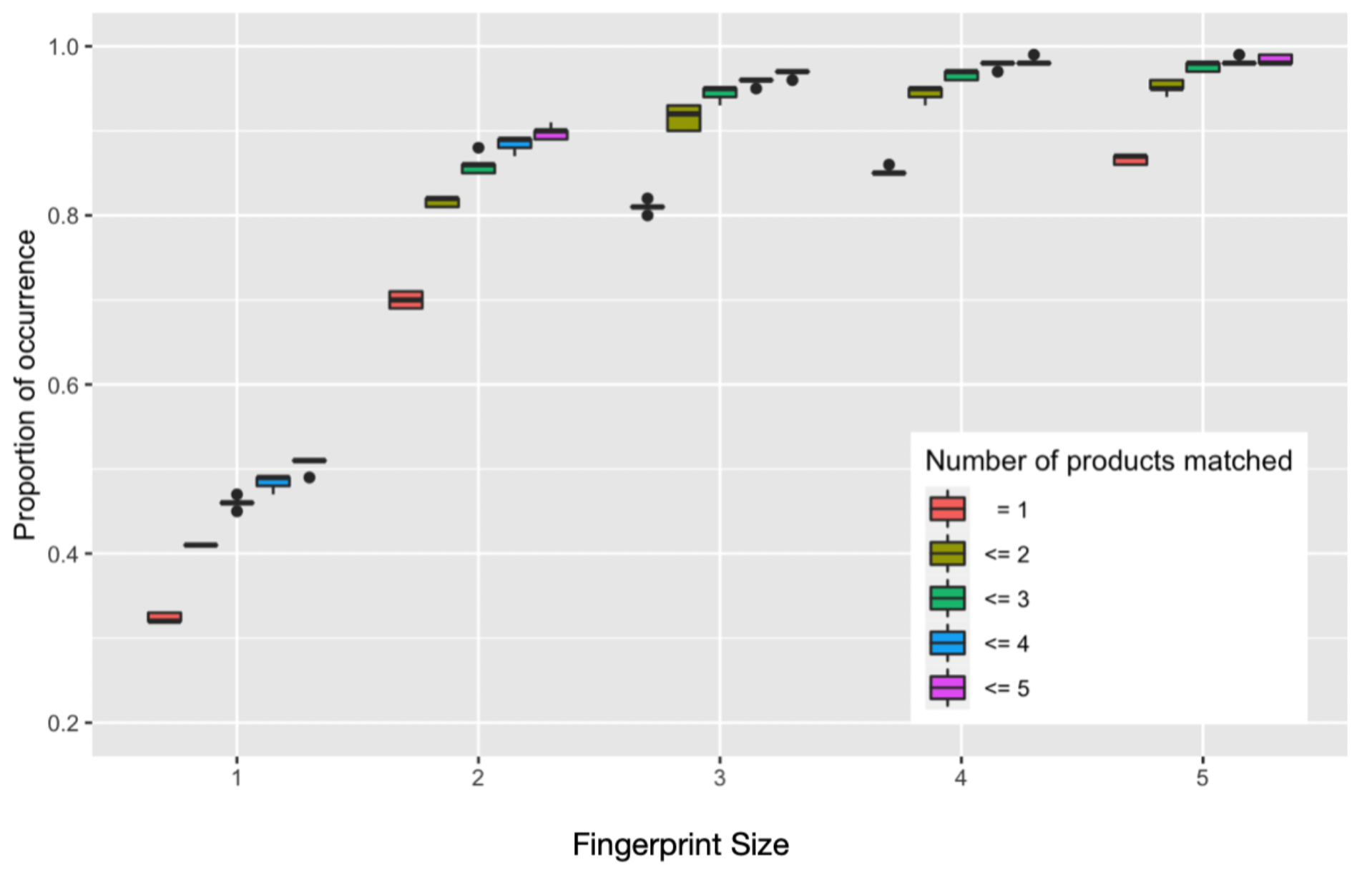}
  \caption{Distribution of the number of candidates: disjoint files strategy
  }
  \label{fig:rq2_c}
\end{figure}

In \Cref{fig:rq2_b,fig:rq2_c} we show the cumulative distributions of the number of candidates selected, respectively, by the single file and disjoint files fingerprinting strategies.
Results are shown for varying \fingerp size from 1 to 5.
To account for randomness in fingerprint selection, each experiment was run 5 times. For each \fingerp size the figures show a box plot of 5 points.
Each box plot value shows the accumulated proportion of searches in which the number of products matched was less of equal to a certain number of different products (from 1 to 5).
For example, in the 5 experiments using single-file sampling and \fingerp size equal to 1, the proportion of times that the
\fingerp matched three or less product was 44.1\%, 46.7\%, 46.9\%, 47.2\%, and 47.5\%; in other words in
a median of 46.9\% cases, the number of products matched was less or equal to 3 (34.6\% returned 1, 7.8\% returned 2,
and 4.5\% returned 3). As it can be seen, in all cases the 5 runs returned very similar results---suggesting that the
method is very stable.

For both strategies, there is a significant reduction in number of matches from a \fingerp of size 1 to 3. After 3 we reach a point of diminishing returns. For a fingerprint of size 3, the single file strategy finds 5 or less matched products in a median of
93.1\% of cases, while the disjoint file strategy finds 5 or less matches in a median of 96.7\% of cases. Equally
important is that with a \fingerp of size 3, 76.7\% of cases returned exactly one match for the single file strategy,
and 81.3\% for the disjoint file one. The disjoint files sampling strategy performs best, likely because sampling from different
files provides more information regarding the project the identifiers belong to.

\begin{hassanbox}{\columnwidth}[4ex]
  When three non-common (after blocklist exclusion) identifiers are sampled from different files of a project, the
  number of projects with those identifiers is 1 in 80\% of cases (precision 100\%), and at
  most 5 in 98\% of cases (precision 20\%). For the single-file method, the number of
  results of size 1 is in 74\% of cases, and at most 5 in 89.2\% (the recall is always 100\%).
\end{hassanbox}

 \section{Evaluation}
\label{sec:validation}

In this section we empirically evaluate the effectiveness of our approach for
software provenance identification based on identifiers. To that end, we use as
subjects a set of software packages that were not extracted directly from our
\pypi corpus (compared to what we did in \Cref{sec:rq2}), but that we expect to be present in
this corpus.  Specifically, we consider version 10 ``Buster'' of the Debian
GNU/Linux distribution and assume that Python software shipped by Debian is
also present in \pypi. This is a reasonable assumption, because Debian is fairly
selective in what it ships in its stable releases and \pypi is the most
comprehensive listing of Python packages. Thus, we expect \pypi to be a
\emph{superset} of the Python packages shipped in Debian stable release like
Buster and, conversely, Debian Python packages to be a ``\emph{golden
  (sub)set}'' of \pypi.

\subsection{Creating a golden set from Debian}

We started by listing all Debian source packages shipped by Debian Buster as of
August 2020,\footnote{At that time Debian Buster was already shipped as a
  ``stable'' release, so while it is possible that its content has changed
  since, modifications are expected to be minimal according to Debian release
  processes.} at the same time we built our \pypi corpus (see
\Cref{sec:rq1method}). The initial list includes over \num{28000} Debian source
packages. We first determined the list of Python packages using a coarse
heuristic based on package names: we selected the source packages that contain
any of the following substrings in either their name or in the name of any of
the binary packages they generate: \texttt{"py"} and \texttt{"python"}.
We identified \num{3155} Debian source packages related to Python this way. We
downloaded the full source code of each package running the command
\texttt{apt-get source \emph{PACKAGE\_NAME}} on a Debian Buster machine.

One major challenge to create a ground truth is to match a package in Debian to
its corresponding package in \pypi, as they might have different names. \pypi
requires packages to include a \texttt{setup.py} file which, in its implementation, should
eventually call the \verb|distutils.setup()| function passing a \texttt{name}
parameter that matches the \pypi package name. We used two best-effort methods
to identify \pypi product names based on this.

First, we ran \verb|python setup.py --name|, with both versions 2 and 3 of the
Python interpreter (because Debian source packages can be implemented in either
version of the language, and there are subtle syntactic differences between the
two language versions). If this script failed, we inspected manually the package to find its \pypi origin. We were able
to match \num{2221} Debian package names to \pypi projects in our corpus. 40 packages did not have 
any identifiers and were ignored (e.g., the Debian package \verb@python-xstatic-ds3@
originates from the \pypi
package \verb@xstatic-ds3@, but it only contains \verb@__init__.py@ and \verb@setup.py@ files that do not declare any
global identifier).

In the end, our golden dataset was composed of \num{2181} pairs. Each pair was a Debian source
package and its corresponding \pypi project.

\subsection{Ranking candidates}

Our provenance detection method returns a set of candidates for further
inspection. Ideally we would like this set to be a singleton. However, we
expect that more than one package will be returned in the general case.
Thus, it will be beneficial to rank the candidates in
such a manner that the most likely candidates appear first. This way candidates
with a higher probability of being the correct answer will be examined sooner.

Document term frequency is used in several methods for ranking result
candidates (such as tf-idf-weighting). Unfortunately it is not applicable for this purpose because if an identifier is declared two or more times
in package A more than in package B, A is still equally relevant as a source of the identifier than package B. Other
methods rely on partial matches to rank the results; these methods cannot be used 
in our experiments because
we are interested only in documents that match all identifiers in the fingerprint.

Instead, for ranking results we leverage SourceRank~\cite{saini2020popmetrics}, a
well-established ranking metric for open source software packages developed
by the Libraries.io
project.\footnote{\url{https://docs.libraries.io/overview.html\#sourcerank},
  accessed 2021-12-07} Roughly speaking, SourceRank is a compound metric that
takes into account both package \emph{popularity} (based on GitHub ``stars'',
for example) and several \emph{quality} metrics such
as the presence of metadata, license information, README, etc.

We conjecture that if a package is popular (as SourceRank indicates), it is more likely to be the
origin of an entity than one that is less popular.
In these
evaluation experiments, when a set of $N$ candidate \pypi packages is returned,
we then order them by their SourceRank in descending order, thus returning a list
(where position 1 is the package with the highest Source Rank, i.e., the most relevant
package).

\subsection{Methodology}

In this experiment we want to evaluate the \emph{single-file}  and \emph{disjoint-files} strategies for provenance
identification with a \fingerp of size 3 and a \bl of 300.
Because these methods are randomized, we perform 5 trials for each of the strategies.

For the single-file strategy, one file is sampled per trial. From this file, 3 identifiers (not in the blocklist) are
randomly sampled.

For the disjoint-files strategy, in each trial 3 different files are randomly sampled (without replacement), and
from each file one identifier (not in the blocklist) is randomly sampled. At each trial level, all source code files are considered
(thus, it is possible that two or more trials use the same files).

Except for the randomly extracted identifiers, no other information is used in these trials.
To evaluate the effectiveness of the proposed method we computed the following
metrics:
\begin{itemize}

\item The size of the candidate set $C$ obtained with each sampling
  strategy. The smaller, the better: a size of 1 is ideal.

\item Given that our results are ranked, we computed the recall and precision for the top k-results. In general, as k
  increases, precision is expected to drop and recall to increase. For this method to be successful we expect that, for
  small values of k both precision and recall to be high.
\end{itemize}

\subsection{Results}

\Cref{tab:outcomes} summarizes the obtained evaluation results for the single-file and disjoint-files strategies.  As it
can be seen, a number of outcomes did not yield any candidate (no file was found in our corpus): 5.8\% for single-file
method, and 11.0\% for the disjoint method; we will revisit this result in Section~\ref{sec:falseNeg}.

The overall recall of the experiments (i.e.  the number of outcomes that included the correct origin of the product) was
93\% for single-file and 87.6\% for disjoint-files. The proportion of outcomes that yielded at least one candidate, but
none was the true origin of the package, was 1.2\% and 1.3\% for each method, respectively.

\begin{table*}
  \caption{Overall results when matching Debian python packages to the \pypi corpus.}
  \centering \begin{tabular}{l|rr|r}
    \toprule
                               & {Single-file}    & Disjoint-files \\
    \hline
    Subjects                   & \num{2147}             & \num{1631}           \\
    \hline
    Outcomes                   & \num{10735}            & \num{8155}           \\
    \quad Empty (no candidates)                & 619 (\ 5.8\%)      & 900 (11.0\%)   \\
    \quad Non-empty candidates    &                  &                 \\
    \quad \quad Successful (i.e. \textbf{overall recall})    & 9982    (93.0\%) & \num{7145} (87.6\%)  \\
    \quad \quad Failed         & 134   (\ 1.2\%)    & 110 (\ 1.3\%)    \\
    \hline
    Size of result per outcome &                  &                 \\
    \quad Median               & 1                & 1              \\
    \quad Avg                  & 14.5             & 2.4            \\
    \quad Max                  & \num{2104}           & 175            \\
  \bottomrule
\end{tabular}
  \label{tab:outcomes}
\end{table*}

We computed the precision, recall and F-score of the top
k-results. This method of using precision and recall is widely used to evaluate the quality of
search results. For a specific search, its precision at k is
defined as number of relevant items in the first k-results; while recall is defined (as usual) as the number of
retrieved items in the first k-results divided by the number of all possible relevant items.

In our experiment there is only one relevant item for any search, therefore the search has a binary outcome: either
the correct product is part of the results or not.
The recall at k is either
1 if the correct product is found in the first k-results, or zero otherwise.
Therefore, the average recall at k is equal to the proportion of successful outcomes when the subject is found in the
first k results of each search.

For a given search, the precision at k is $1/min(k, n)$ if the correct product is in the first k-results and zero otherwise (where $n$ is the overall number of items in the result of the search).
Because there is only one potential correct item, the precision at k drops rapidly as k increases and the query
returns at least k elements.
We compute the average precision at k 
over all outcomes ignoring
the cases where the result returns zero results (precision is undefined; approximately 10\% of results fell into this
category).

For the sake of readability, for the rest of this paper we will refer to 
average recall at k and average precision at k
simply as recall at k and precision at k.

\begin{figure}
  \centering \includegraphics[width=0.48\textwidth]{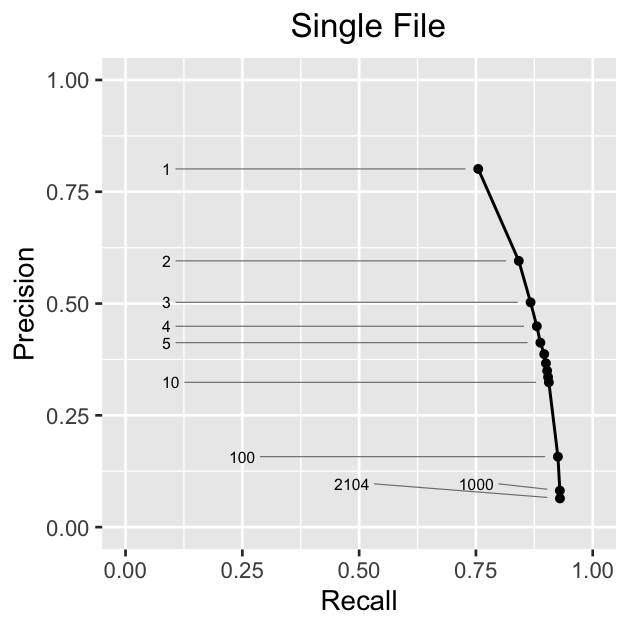}
  \includegraphics[width=0.48\textwidth]{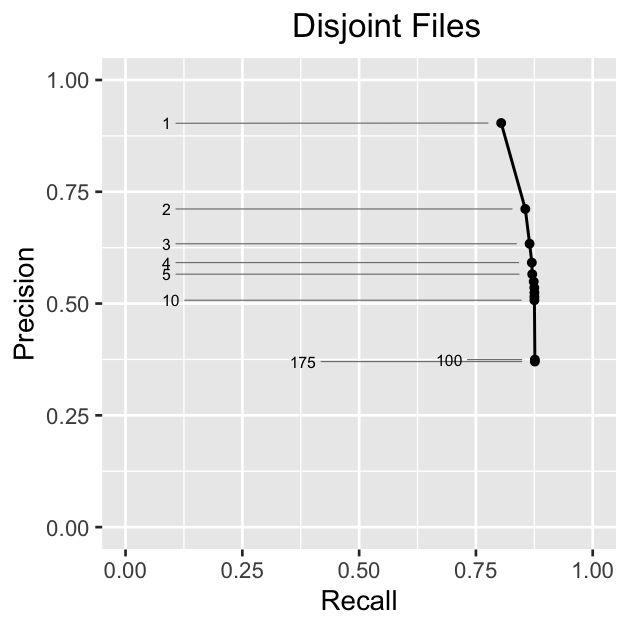}
  \caption{Average recall and precision at k for both strategies.
    The small numbers to the left correspond to k.
    For exampple, for the single-file strategy, the top result (k=1) has a recall of 75\%, and a precision of
    80\%. }
  \label{fig:bothk}
\end{figure}

Figure~\ref{fig:bothk} shows these values for the single-file strategy, which are
also shown in Table~\ref{tab:singlekscore}. As it can be
seen, for $k=1$, the precision is 80\% and the recall is 75\% (this is because 7.6\% of outcomes did not return any
result).
With $k>5$, the gains in recall are very marginal; yet, at $k=10$,
  the precision is almost $1\over 3$. The table also includes the number and proportion of outcomes that yielded a result of
  size $k$ or larger; for example 89.0\% of outcomes returned one or more candidates (31.1\% returned exactly one
  candidate), and 16.4\% returned more than 6 or more candidates.

\begin{table}[ht]
  \caption{Average recall, precision and F-score at k for single-file strategy. The number and
    proportion of relevant outcomes corresponds to those that had at least k candidates; e.g. only 31\% of outcomes had
    3 or more candidates.}
\centering
\begin{tabular}{r|rr|rrrrr|}
  \hline
   k & \multicolumn{2}{c|}{Relevant Outcomes} & Recall & Precision & F-score \\
    & Count & Prop. &\\
  \hline
  1    & 10116 & 94.2\% & 0.75 & 0.80 & 0.78 \\ 
    2  & 5059  & 47.1\% & 0.84 & 0.60 & 0.70 \\ 
    3  & 3337  & 31.1\% & 0.87 & 0.50 & 0.64 \\ 
    4  & 2523  & 23.5\% & 0.88 & 0.45 & 0.59 \\ 
    5  & 2069  & 19.2\% & 0.89 & 0.41 & 0.56 \\ 
    6  & 1756  & 16.4\% & 0.90 & 0.39 & 0.54 \\ 
    7  & 1504  & 14.0\% & 0.90 & 0.37 & 0.52 \\ 
    8  & 1334  & 12.4\% & 0.90 & 0.35 & 0.50 \\ 
    9  & 1228  & 11.4\% & 0.90 & 0.34 & 0.49 \\ 
   10  & 1111  & 10.3\% & 0.91 & 0.32 & 0.48 \\ 
  100  & 169   & 1.6\% & 0.93 & 0.16 & 0.27 \\ 
  1000 & 34    & 0.3\% & 0.93 & 0.08 & 0.15 \\ 
  2104 & 2     & 0.0\% & 0.93 & 0.06 & 0.12 \\ 
   \hline
\end{tabular}
\label{tab:singlekscore}
\end{table}

\begin{table}[ht]
  \caption{Average recall, precision and F-score at k for disjoint-files strategy.}
\centering
\begin{tabular}{r|rr|rrrrr|}
  \hline
   k & \multicolumn{2}{c|}{Relevant Outcomes} & Recall & Precision & F-score \\
    & Count & Prop. &\\
  \hline
    1 & 7255 & 89.0\% & 0.80 & 0.90 & 0.85 \\ 
    2 & 2552 & 31.3\% & 0.86 & 0.71 & 0.78 \\ 
    3 & 1322 & 16.2\% & 0.86 & 0.63 & 0.73 \\ 
    4 & 858  & 10.5\% & 0.87 & 0.59 & 0.70 \\ 
    5 & 563  &  6.9\%  & 0.87 & 0.57 & 0.69 \\ 
    6 & 428  &  5.3\%  & 0.87 & 0.55 & 0.67 \\ 
    7 & 353  &  4.3\%  & 0.87 & 0.54 & 0.66 \\ 
    8 & 282  &  3.5\%  & 0.88 & 0.52 & 0.66 \\ 
    9 & 245  &  3.0\%  & 0.88 & 0.52 & 0.65 \\ 
   10 & 207  &  2.5\%  & 0.88 & 0.51 & 0.64 \\ 
  100 & 11   &  0.1\%  & 0.88 & 0.37 & 0.52 \\ 
  175 & 1    &  0.0\%  & 0.88 & 0.37 & 0.52 \\ 
   \hline
\end{tabular}
\label{tab:diskscore}
\end{table}

For the disjoint-files strategy, the precision and recall at $k$ is shown in Figure~\ref{fig:bothk} and
Table~\ref{tab:diskscore}. For this strategy 11\% of outcomes did not generate any result.
As it can be seen, the recall and precision is very high for $k=1$ (0.8 and 0.9 respectively), and for $k>2$ there is no
significant gain in recall. Thus, in most cases, it suffices to inspect only the top 2 results, with a recall of
86\% and a precision of 71\%.  Inspecting results beyond
the second candidate only yielded a successful result in 2\% of the outcomes. Also, only 16.2\% of all searches return
more than 2 candidates, and 5.3\% more than 6 or more candidates.

Comparing the two methods we observe that:
\begin{itemize}
\item For $k=1$, the disjoint strategy has better recall (0.80 vs 0.75).

\item The single-file strategy returned larger lists of candidates than the disjoint one.
As a consequence, the precision of the single-file strategy drops much more 
  as $k$ increases, 

\item For $k\ge4$, the single-file strategy has better recall than the disjoint strategy.
\end{itemize}

While the precision of the disjoint-files strategy is higher than the single-file one, not all packages had
3 files and after $k>2$ the recall of the single-file strategy is higher.

\begin{hassanbox}{\columnwidth}[4ex]
  Both strategies (single-file and disjoint-files) are effective at finding the origin of a Debian Python package in
  \pypi without the need to inspect many candidates.
  At k=1, the recall and precision of the single-file strategy were 0.75 and 0.8; while for 
  the disjoint-files strategy, they were 0.8 and 0.9, respectively.
\end{hassanbox}

\subsection{False negatives}
\label{sec:falseNeg}

Since we manually curated the dataset we expected to have no false negatives (i.e. we know the Debian package exist in
\pypi), yet the proposed method is unable to identify the origin of several packages.  As shown in table \Cref{tab:outcomes}, the single-file
strategy did not return any results in 5.8\% of the outcomes and in 1.2\% of the outcomes the results did not include
its true origin. For the disjoint-files strategy, the number of outcomes with empty results was almost twice (11.0\%) and
the number of outcomes that did not include its true origin was almost the same (1.3\%).

There were 21 packages that were never matched to a package in \pypi by either strategy.
We manually inspected each of them. We observed that
 the reason these packages resulted in false negatives was that their corresponding packages in \pypi did not contained all the original
source files of the packages (which the Debian package included). We observed two cases:

\begin{itemize}
\item Some \pypi packages did not include testing or examples source code. 19 of the 21 packages did not include Python files located in the  test, examples, or documentation folders.
  Many packages in \pypi were \emph{binary} distributions (called distribution archives by the
  Python Packaging Authority). These distribution packages were a subset of the
  original source code and contain the files necessary to use the package, not to build it
  \cite{packageGuidelinesDebian}. Some of these 19 packages had very few source files in \pypi and many more test source files in Debian. For
  example, \emph{pyfaidx} had 3 Python files (\verb@setup.py, __init__, cli.py@) in \pypi, and 21 in Debian (the other 18 files
  were inside the test folder).
\item Some \pypi packages contained only installation scripts.
The remaining two packages (\emph{egenix-mx-base} and \emph{QuantLib}) only included installation scripts in \pypi without
  most of its source code. For example, \emph{egenix-mx-base} had 85 files in Debian and only 2 in \pypi. In the case of
  \emph{QuantLib}, the numbers were 23 and 1 respectively.
\end{itemize}

Given the first point above, we hypothesized that false positive outcomes would include files in \texttt{test} of
\texttt{example} paths and that Debian packages would be more likely to include files in \texttt{test} or
\texttt{example} folders.
We checked each of the 
false negatives outcomes to see if the sample file had \texttt{test} or \texttt{example} in the name of a folder in the
path of the file (checking if the full filename matched the regular expression \verb@(text|example).*/)@); for disjoint-files strategy we check if at least one file in the outcome satisfied this condition). The
results are presented in Table~\ref{tab:falsePositives}. As it can be seen, for outcomes that had an empty result (no
candidates), 75\% of files were in test or example folders for the single-file strategy; for the disjoint-files strategy
it was 89.2\%. In total, 70\% of the false negative outcomes in the single-file strategy were in folders named test or
example, and for the disjoint-files strategy they were 87\%.

We checked the number of packages with tests or example files in our experiment. Out of the 2147 Debian packages used in
our experiment, 1461
had at least one of these types of files. In contrast, of the corresponding 2147 \pypi packages, 1461 had one or more
test of example files. Therefore, the
problem of missing test files seems to be restricted to a small proportion of packages. In fact, in the
single-file strategy experiment, 3,198 trials that used a test or example file were successful (out of 3,709, 86.2\%).
Thus test and example files contain identifiers that are useful
for provenance discovery.

\begin{table*}
  \caption{Most of the false negatives used one or more files placed in folders called \texttt{test} or
    \texttt{example}. We manually inspected 19 out of 21 that only had false positives
    and discovered that such files were present in the Debian package and its source code repository,
    but not in the \pypi packages.}
  \centering \begin{tabular}{l|rr|rr}
    \toprule
                               & \multicolumn{2}{c|}{Single-file}    & \multicolumn{2}{c}{Disjoint-files} \\
    \hline
    Outcomes                   &   753  &          &  1010  &        \\
    \quad Had file(s) in test or example folder & 525 & 69.8\%           &  880 & 87.1\% \\
    \hline
    No candidates             & 619   &           & 900 & \\
    \quad Had file(s) in test or example folder & 456 & 73.7\%           &  803 & 89.2\% \\
    \hline
    Non-empty candidates       & 134          &    & 110 &\\
    \quad Had file(s) in test or example folder & 69 & 51.5\%           &  77 & 70.0\% \\
  \bottomrule
\end{tabular}
  \label{tab:falsePositives}
\end{table*}

\begin{table*}
  \caption{Results for best outcome (out of 5 trials per project) for the \emph{single-file} sampling strategy: \num{2147}
    subjects; and for disjoint-files: 1631 subjects.}
  \centering \begin{tabular}{l|rr|r}
    \toprule
                               & {Single-file}    & Disjoint-files \\
    \hline
    Subjects                   & 2147             & 1631           \\
    \hline
    Outcomes                   & 2147            & 1631           \\
    \quad Empty (no candidates)                & 38 (\ 1.8\%)      & 116 (7.1\%)   \\
    \quad Non-empty candidates    &                  &                 \\
    \quad \quad Successful (\textbf{overall recall})    & 2099    (97.8\%) & 1501 (92.0\%)  \\
    \quad \quad Failed         & 10   (\ 0.5\%)    & 14 (\ 0.9\%)    \\
    \hline
    Size of result per outcome &                  &                 \\
    \quad Median               & 1                & 1              \\
    \quad Avg                  & 4.5             & 2.2            \\
    \quad Max                  & {2031}           & 175            \\
  \bottomrule
\end{tabular}
  \label{tab:outcomes-best}
\end{table*}

These results imply that our experiments yielded false positives and empty results
because, when a file is picked at random, the chosen file is not included in the \pypi
package (such as the files in \texttt{test} and \texttt{example} folders as described above). In the design of our experiment we 
took into account some of the effects of this
randomness, and repeated the search 5 times per project. Table~\ref{tab:outcomes-best} shows the summary of the results
when we pick the best outcome per project. When these results are compared with Table~\ref{tab:outcomes} we can observe
that the overall recall jumps from 93.0\% to 97.8\% for the single-file strategy, and from 87.6\% to 92.0\% for the
disjoint-files strategy; the best outcome did not find the origin of a package in 2.2\% (48) projects for
the single-file strategy and 8\% (130) for the disjoint-strategy.

\begin{table}[ht]
\caption{Best outcome (out of 5 trials per project), single-file strategy.}
\centering
\begin{tabular}{r|rr|rrrrr|}
  \hline
   k & \multicolumn{2}{c|}{Relevant Outcomes} & Recall & Precision & F-score \\
    & Count & Prop. &\\
  \hline
    1 & 2109 & 98.23 & 0.90 & 0.92 & 0.91 \\ 
    2 & 924 & 43.04 & 0.95 & 0.67 & 0.79 \\ 
    3 & 559 & 26.04 & 0.96 & 0.57 & 0.72 \\ 
    4 & 388 & 18.07 & 0.96 & 0.52 & 0.68 \\ 
    5 & 290 & 13.51 & 0.97 & 0.49 & 0.65 \\ 
    6 & 218 & 10.15 & 0.97 & 0.46 & 0.63 \\ 
    7 & 177 & 8.24 & 0.97 & 0.45 & 0.61 \\ 
    8 & 140 & 6.52 & 0.97 & 0.43 & 0.60 \\ 
    9 & 123 & 5.73 & 0.97 & 0.42 & 0.59 \\ 
   10 & 101 & 4.70 & 0.97 & 0.41 & 0.58 \\ 
  100 &   5 & 0.23 & 0.98 & 0.30 & 0.46 \\ 
  1000 &   1 & 0.05 & 0.98 & 0.25 & 0.39 \\ 
  2031 &   1 & 0.05 & 0.98 & 0.22 & 0.36 \\ 
   \hline
\end{tabular}
\label{tab:dis-score-best}
\end{table}

\begin{table}[ht]
\caption{Best outcome (out of 5), disjoint-files strategy.}
\centering
\begin{tabular}{r|rr|rrrrr|}
  \hline
   k & \multicolumn{2}{c|}{Relevant Outcomes} & Recall & Precision & F-score \\
    & Count & Prop. &\\
  \hline
    1 & 1515 & 92.89 & 0.88 & 0.95 & 0.92 \\ 
    2 & 494 & 30.29 & 0.91 & 0.74 & 0.82 \\ 
    3 & 245 & 15.02 & 0.91 & 0.66 & 0.77 \\ 
    4 & 150 & 9.20 & 0.92 & 0.62 & 0.74 \\ 
    5 &  94 & 5.76 & 0.92 & 0.60 & 0.72 \\ 
    6 &  66 & 4.05 & 0.92 & 0.59 & 0.72 \\ 
    7 &  56 & 3.43 & 0.92 & 0.57 & 0.71 \\ 
    8 &  40 & 2.45 & 0.92 & 0.56 & 0.70 \\ 
    9 &  38 & 2.33 & 0.92 & 0.56 & 0.69 \\ 
   10 &  32 & 1.96 & 0.92 & 0.55 & 0.69 \\ 
  100 &   1 & 0.06 & 0.92 & 0.41 & 0.57 \\ 
  123 &   1 & 0.06 & 0.92 & 0.41 & 0.57 \\ 
   \hline
\end{tabular}
\label{tab:disk-score-best}
\end{table}

Table~\ref{tab:dis-score-best} shows the recall and precision at k for best outcome for single-file strategy; and
Table~\ref{tab:disk-score-best} for best outcome for disjoint-files strategy (out of 5 trials in both cases). As it can be seen, for k=1, the F-score of single-file strategy
is 0.91, and 0.92 for the disjoint-files strategy. 
Note that the numbers for disjoint-files strategy are (with the exception of precision and F-score at k=1) worse than the
single-file strategy, implying that, when we take the best of various trials, it is best to use the single-file strategy.

\begin{hassanbox}{\columnwidth}[4ex]
Manual inspection of the false positives appears to indicate that several packages in \pypi do not have all the source
code of their corresponding packages. This problem can be alleviated by repeating the search several times. In our
experiments, when using the best of 5 trials result, the average precision and recall at k=1 of the single-file method improves to 0.90 and
0.92, and to 0.88 and 0.95 for the disjoint-files method (i.e. when only inspecting the top result).
\end{hassanbox}

\subsection{An improved algorithm for provenance discovery using identifiers}

Overall, these results suggest the following algorithm that will only inspect 5 candidates at most, and uses the
single-file strategy:

\begin{minipage}[c]{1.0\linewidth}
\begin{itemize}
\item Repeat at most 5 times: 
\begin{enumerate}
\item Pick one file at random with replacement.
\item From this file, extract 3 random identifiers not in the blocklist.
\item Search the corpus for candidates.
\item Inspect only the top-candidate:\\ If it is the origin of the package stop.
\end{enumerate}
\end{itemize}
\vspace{3mm}
\end{minipage}

\textbf{In our experiments, this algorithm has a recall of 0.90 and precision of 0.77} (it would have required to inspect 
2731 candidates, of which 2099 were correct). Equally important, it would have been
applicable to 30\% more packages than the disjoint method. The processing time of querying an identifier
is negligible, thus most of the CPU time will be consumed extracting the random identifiers from a given file.
The inspection of each top candidate can be assisted with a clone detector that compares the subject against the candidate.

\begin{hassanbox}{\columnwidth}[4ex]
  Using the single-file strategy and repeating the search at most 5 times and only inspecting the top result
  has a recall of 0.90 and precision of 0.77.
\end{hassanbox}

\subsection{Low precision in few outcomes}
\label{sec:lowPrecision}

Most outcomes had a very small set of candidates (the median was 1). However, in some cases the number of candidates was
very large. We manually inspected the results of outcomes with the most candidates. What we observed is that there is a
significant amount of cloning in \pypi packages.
We queried the \pypi corpus to identify packages that had a large number of common identifiers and manually inspected
several of them (further research should conduct a proper study of the existence and frequency of these common identifiers in \pypi).
We identified the following reasons why some identifiers are used in different packages:

\begin{enumerate}
\item Commonly used identifiers. This case corresponds to functions/classes that are frequently used, yet have very different
  source code. We have already discussed them in the creation of the blocklist (see \Cref{sec:freqidents}).
\item Different variants of the same package. Some packages are specialized versions of others. These packages appear to
  be different binary distributions of the same source code, yet each appears as a different \pypi package.
  For example, the
  following packages share the majority of their code with
  \emph{tensorflow}: \emph{tensorflow-cpu, tensorflow-directml, tensorflow-fedora20, tensorflow-gpu,
    tensorflow-gpu-macos, tensorflow-rocm, tensorflow-rocm-enhanced, tensorflow-tflex}; these packages share between
  3801 and 11,802 functions with \emph{tensorflow}.
\item Embedding dependencies. We found that a significant number of packages embed their
  dependencies. Usually this is done during the build process, where the dependencies are located in a
  folder named \texttt{thirparty}. We identified 131 packages that have python files in the folder \texttt{thirparty} and 112 in \texttt{third\_party}.
  For example, \emph{sqlmap} had twenty different packages under \texttt{thirparty}; these dependencies did not
  have any information that documented the version or origin of each of them. 
\item Embedding a dependency into its own source code.
  In this case, the dependency is copied inside the source code tree of the package (usually inside the folder \textsf{src})
  and becomes part of the source code of the package. For example, \emph{nplab} embedded the project \emph{lucam} (a single Python file).
\item Copied functions. Sometimes code is cloned among different projects.
For example, we found one common class (\texttt{ColorizingStreamHandler}) in \emph{sqlmap} and \emph{Mopidy}
  that originated from a gist in github. Both packages properly attributed the origin of this class. In another case,
  the packages  \emph{imgserve} and \emph{Amara} shared only one identical function
  (\texttt{get\_\-filename\_\-parts\_\-from\_url}) with no attribution.
\item Same package, different names. We found one package that was uploaded under different names by different
  maintainers (\emph{hd-llz} and \emph{hand-detector-test}). Furthermore, the source code of
the packages was slightly different (one had 940 files, the other 895).

\item Subclassing. Some libraries expect to be reused via subclassing and dynamic dispatch. In this case,
  the new code will reuse the same identifiers for some of the method's functions. For example, the \emph{meta-blocks}
  defines a class \texttt{MetaBlocksExperimentSearchPathPlugin} derived from the class \texttt{SearchPathPlugin}
 in package \emph{hydra-core}. In this class, \emph{meta-blocks} redefines the method \texttt{manipulate\_search\_path},
 that is originally defined in \emph{hydra-core}.
\item Code generation. Some packages shared names of classes and functions that had been generated by the same tool. For example, \emph{yuuki-core},
  \emph{AsyncLine} and \emph{LineService}  had several classes in common, all in files that included
  the header \texttt{Autogenerated by Thrift Compiler}.
\end{enumerate}

These points emphasize the difficulties of curating a corpus for provenance discovery. During the creation of the
corpus, provenance analysis of each package could be done in order to identify cases such as the ones above towards the
goal
of improving the quality of the corpus.

 \section{Discussion}
\label{sec:discussion}

Developers of \pypi products seem to, consciously or not, choose identifiers that are either unique or very distinctive throughout the entire \pypi ecosystem.
Out of 11.2 million different identifiers for classes and functions/methods (in 244k different products), 76\% are unique (and 95\% appear in at most 4 different products).
Equally remarkable is that the intersection of the names of classes and functions was negligible (less than 0.6\%).

Python has a strict module namespace mechanism.
By default, the user of a library should prefix any of its identifiers with the name of the library (e.g., \verb@pandas.DataFrame@); this implies that
identifiers only need to be unique within the project. At the same time there is an
implicit expectation grounded in Python practices and coding guidelines that developers will
import library identifiers in such a way  that they do not require the full qualified name (e.g., \texttt{from pandas import
DataFrame}) and this might be a motivating reason why identifiers tend to be
unique. Nonetheless, universally unique identifiers seem to also naturally emerge in the corpus.

The median length of identifiers in \pypi is 16 characters for classes and 19 for function names.
This means that developers of \pypi packages are willing to name their identifiers with descriptive names.
Future research should look into the composition of identifiers in terms of its components (e.g., by splitting at underscores or case changes) and abbreviations used.
Filenames, which correspond to module names in Python, are also quite long with
a length of 15 characters.

\begin{figure}
  \centering
  \includegraphics[width=0.9\textwidth]{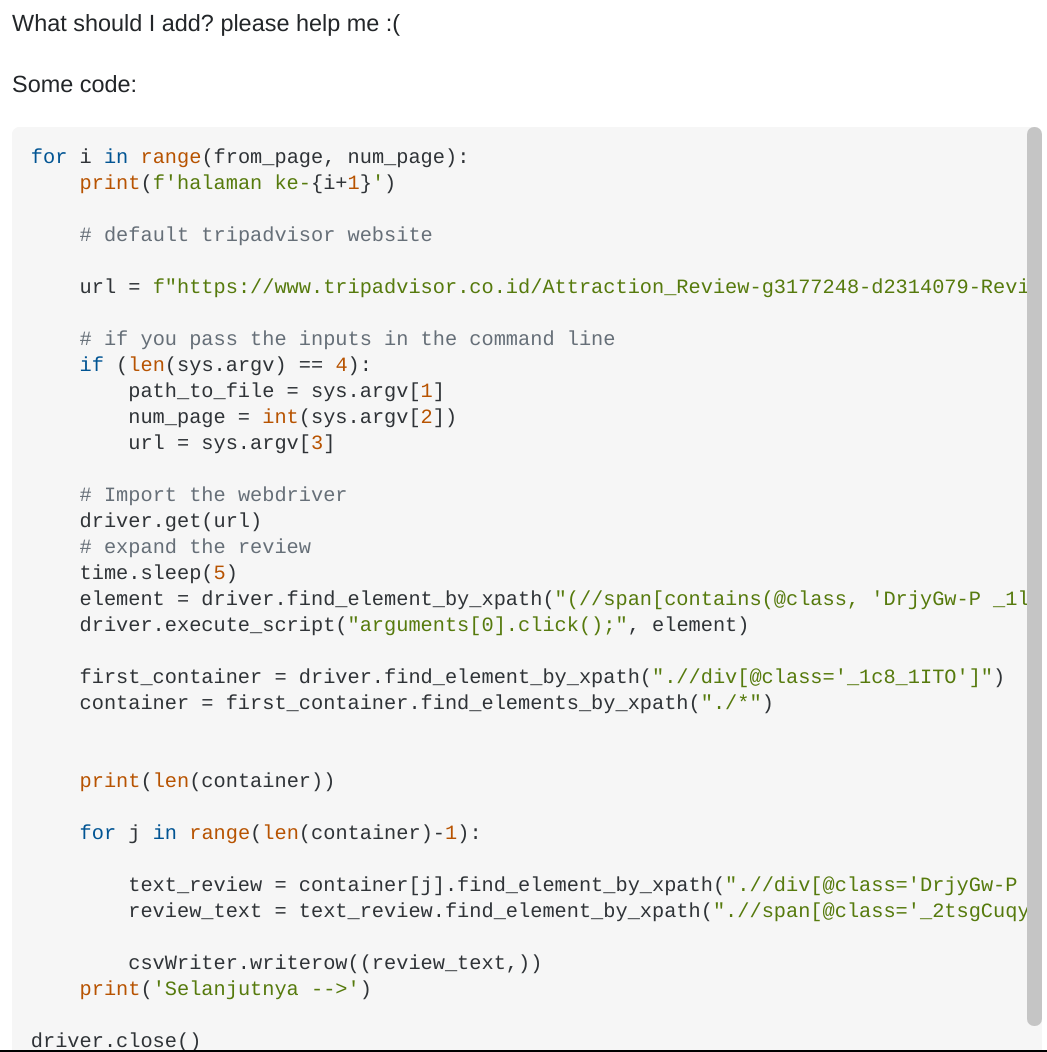}
  \caption{Example of a Stack Overflow question that does not indicate which Python library it is using.
    The identifiers used in it narrow it to 3 candidate: \texttt{selene}, \texttt{selene-kentastik} and \texttt{django-cloud-reploy}, of which \texttt{selene} is the highest ranked by Libraries.io.}
  \label{fig:stackoverflow-post}
\end{figure}

The fact that most identifiers are unique to a package can be leveraged for applications other than software provenance identification.
For instance, it might be possible to scan a Python source code snippet and determine with little or no ambiguity the libraries it uses, even when \texttt{import} statements are omitted.
Consider questions on the Stack Overflow Q\&A website, where only very popular
libraries (such as pandas, numpy, etc.) tend to have dedicated classification
tags on the website;
it is not trivial to identify postings that use other libraries, unless library names are explicitly mentioned in the question title.
Identifiers can be automatically extracted from the code snippet and matched to the corresponding packages.
Take for example Stack Overflow question n.~\num{68878857}\footnote{Titled: ``Scraping tripadvisor review, len container change, no such
  element Unable to locate element'', \url{https://stackoverflow.com/questions/68878857}, accessed 2022-01-16}
depicted in \Cref{fig:stackoverflow-post}, which does not mention which library the snippet is using.
The identifiers \verb+find_element_by_xpath+, \texttt{get},
\verb+execute_script+, and the filename \verb+driver.py+ only exist in 3
different packages (\emph{django-cloud-deploy}, \emph{selene}, and \emph{selene-kentastik}), \emph{selene} (Python
bindings for selenium) being the highest ranked package among them according to
SourceRank (\emph{selene-kentastic} is a different set of selenium bindings for Python---thus
using some identifiers with \emph{selene}).

The Debian experiment demonstrated the effectiveness of a method of provenance discovery that uses
identifiers. Specifically, that by simply conducting at most five trials, and only inspecting the first result returned,
the single-file strategy can achieve a recall of 0.9 and precision of 0.77.

We must emphasize that the effectiveness of any method of software provenance identification depends primarily on how comprehensive the corpus is, and the corpus needs to be maintained current as time passes by.
New product versions tend to keep most previous identifiers, can introduce new ones, but can also remove some.
Thus using \emph{only} global identifiers (and in particular with methods that rely on random sampling, as in our Debian
experiment) is insufficient to correctly pinpoint the \emph{release} of a product. However, once a product has been
identified, it is possible to use set distance metrics (such as Jaccard) or clone detection methods to determine the best candidate release
that matches the candidate.

There are also other potential uses for a corpus of identifiers. For example, it might be possible to identify products
that have evolved from others (i.e., forks) or that have changed name. It might also be possible to identify functions
and classes that are copied from one product to another. Finally, 
IDEs can benefit from such a corpus of identifiers to suggest (or automatically
add) include/import statements.

The goal of the propose approach to provenance identification is not to replace
clone detection tools. On the contrary, the goal is to narrow the potential
search space such that more time consuming methods (including, but not limited
to clone detection ones) can be applied more selectively.

 \subsection{Threats to validity}
\label{sec:threats}

\paragraph{Construct validity.}

We have conducted several integrity tests to verify that our processing is accurate.
We entirely relied on Universal Ctags for the extraction of classes and
function names in Python, hence we depend on its reliability. At the same time it is the state-of-the-art open source tool for identifier indexing and is actively maintained, so we are confident in its quality.

We note that we only processed the most recent 100 releases of each \pypi product.
However, the impact of this decision is probably negligible, since we expect any recent releases to have a large identifier overlap with older releases (we have observed that most releases are a superset of previous releases, only rarely identifiers are removed).

\paragraph{Internal validity.}

In our experiments we assume that \pypi is a trustworthy reference corpus for Python products. However, \pypi relies on
the developers to make sure they upload correct code. We observed that in some cases, developers might upload releases
with embedded dependencies, a phenomenon known as ``dependency vendoring''~\cite{zimmermann2020vendoring}.
This is often addressed in future releases by removing from them the source code of previously embedded dependencies, but the older bloated releases remain in \pypi.
Since Python does not have a way to restrict the visibility of an identifier, any global identifier in a product is thus
available outside it. It is very likely that developers use different naming mechanisms for identifiers that are
expected to be used by others from those that they are only to be used locally.

\paragraph{External validity.}

Our results only apply to the ecosystem of packages in \pypi and written in Python.
We do not claim that these results apply to other programming languages or ecosystems.
In fact, we believe that an empirical evaluation similar to the one we have conducted in this work should be conducted on each major programming language ecosystems, to document and compare the level of uniqueness/distinctiveness across languages.
It would also be interesting to analyze identifier distinctiveness when considering multiple programming languages \emph{together}.
While there are identifiers that will be expected to be shared across languages (such as the case of language bindings that we have identified and discussed) it is possible that identifiers will be fairly unique across languages, helping with software provenance tracking in contexts where the programming language of the code under audit has not been determined, for whatever reason.

 \section{Conclusions}
\label{sec:conclusion}

In this paper we have determined that in the Python ecosystem of libraries approximately 75\% of global identifiers
are unique. Furthermore we have also identified a set of identifiers that are too common to use, and therefore, can be
considered as ``stop words'' in identifier analyses.

We have then used this property to developed a randomized approach to identify the provenance of a
Python library that uses a very small set of globally defined identifiers to
identify the origin of a product. 
The approach is straightforward to implement, fast, and had a recall of 0.9 and precision of 0.77.
experiments, when inspecting only the top result in each of the five trials for the single-file strategy.
In spite of its high accuracy the proposed approach is not meant to be used in isolation, but rather as a preliminary filtering step before applying more expensive identification techniques (if and when multiple candidates are identified), such as code clone detection.

\subsection{Future work}
\label{sec:futurework}

Several research directions, including several empirical studies, remain to be pursued as future work.
A promising one is further increasing the granularity of detection, reaching down to the level of code snippets--such as individual functions or classes extracted from complete source code files, or snippets posted in isolation on social coding websites and platforms.
In principle, the proposed approach is completely agnostic to granularity: it would work at the level of snippets as it does as that of products and releases.
In practical terms, however, two challenges exist: 1) when an entity is duplicated in the corpus, which copy should be
considered the canonical origin, and 2) accuracy would be different at the snippet level, although it is not clear if for the better or worse.
A large-scale empirical experiment is needed, either by splitting up functions/classes from corpora like the one we have already investigated in this work (\pypi) or relying on snippet-first datasets such as GitHub gists or Stack Overflow snippets.

The proposed approach is also agnostic to programming languages. The only requirement being the ability to create a
comprehensive corpus of identifiers for the desired programming language.
Large-scale empirical experiments targeting different programming languages and/or package ecosystems are needed to verify if the language independence of the model translates to good accuracy in other contexts.
It is possible that programming convention in different communities would result in different levels of  identifier uniqueness that could in turn impact accuracy, for better or worse.

The potential synergies between the introduced method and traditional clone detection techniques also deserves further exploration.
With few exceptions, scaling clone detection to large software repositories remains an open challenge. This is
particularly true when
one considers that, for provenance discovery, clones of a source code snippet are not necessarily copies, thus
increasing --from the point of view of provenance discovery-- the number of false positives.
Future work should evaluate a hybrid approach, where identifiers are used to narrow the potential number of candidates, and then
using clone detection tools to finally identify the provenance of source code. Also, future work should explore methods
to improve the qualify of a corpus (such as removing instances of copies of dependencies or identifying variants of the
same product). This is another area where such a hybrid approach can help.

\section*{Data availability}

A replication package for this paper is available at \url{https://doi.org/10.5281/zenodo.7637703}
\cite{sun_yiming_2023_7637703}. This replication package contains all pairs (identifier, product)  found in \pypi.

\end{document}